\RequirePackage{fix-cm}

\documentclass[twocolumn,epjc3]{svjour3}  

\smartqed  
\RequirePackage{graphicx}
\usepackage{lineno}
\usepackage[separate-uncertainty=true,per-mode=symbol]{siunitx}
\usepackage{url}
\usepackage{isotope}
\usepackage{placeins}

\usepackage[colorlinks=true,citecolor=blue,filecolor=blue,linkcolor=blue,urlcolor=blue,pdftex]{hyperref}
\usepackage{currfile} 
\usepackage{orcidlink} 

\DeclareSIUnit\clight{\text{\ensuremath{c}}}
\DeclareSIUnit\gevm{\GeV\per\clight\squared}
\DeclareSIUnit\years{y}
\DeclareSIUnit\tonne{t}
\DeclareSIUnit\tonneyears{\tonne\years}
\DeclareSIUnit\cmtwo{\cm\squared}
\DeclareSIUnit\kev{\kilo\eV} 
\DeclareSIUnit\mev{\mega\eV} 
\DeclareSIUnit{\GeV}{\giga\eV}
\DeclareSIUnit\kevnr{\si{\kev} {}_\mathrm{NR}} 
\DeclareSIUnit\PE{\mathrm{PE}}
\DeclareSIUnit\mus{\micro\second}
\DeclareSIUnit\ns{\nano\second}

\newcommand{\tapprox}{\raisebox{0.5ex}{\texttildelow}}

\newcommand{\hydrogencaptureenergy}{\qty{2.22}{\mev}}
\newcommand{\ambenone}{\qty{4.44}{\mev}}

\newcommand{\amberate}{\qty{159\pm4}{neutrons \per s}}
\newcommand{\ambenrpurity}{$>$\qty{99.9}{\percent}}
\newcommand{\neutroncapturetimewater}{$\tapprox$\qty{200}{\micro \second}}

\newcommand{\taggingefficiency}{\qty{53\pm3}{\percent}}

\journalname{Eur. Phys. J. C}
\begin{document}

\title{The neutron veto of the XENONnT experiment: Results with demineralized water
}

\author{E.~Aprile\orcidlink{0000-0001-6595-7098}\thanksref{addr3}
\and
J.~Aalbers\orcidlink{0000-0003-0030-0030}\thanksref{addr28}
\and
K.~Abe\orcidlink{0009-0000-9620-788X}\thanksref{addr23}
\and
S.~Ahmed Maouloud\orcidlink{0000-0002-0844-4576}\thanksref{addr18}
\and
L.~Althueser\orcidlink{0000-0002-5468-4298}\thanksref{addr7}
\and
B.~Andrieu\orcidlink{0009-0002-6485-4163}\thanksref{addr18}
\and
E.~Angelino\orcidlink{0000-0002-6695-4355}\thanksref{addr14,addr4}
\and
D.~Ant\'on~Martin\orcidlink{0000-0001-7725-5552}\thanksref{addr1}
\and
F.~Arneodo\orcidlink{0000-0002-1061-0510}\thanksref{addr9}
\and
L.~Baudis\orcidlink{0000-0003-4710-1768}\thanksref{addr17}
\and
M.~Bazyk\orcidlink{0009-0000-7986-153X}\thanksref{addr13}
\and
L.~Bellagamba\orcidlink{0000-0001-7098-9393}\thanksref{addr0}
\and
R.~Biondi\orcidlink{0000-0002-6622-8740}\thanksref{addr6}
\and
A.~Bismark\orcidlink{0000-0002-0574-4303}\thanksref{addr17}
\and
K.~Boese\orcidlink{0009-0007-0662-0920}\thanksref{addr6}
\and
A.~Brown\orcidlink{0000-0002-1623-8086}\thanksref{addr19}
\and
G.~Bruno\orcidlink{0000-0001-9005-2821}\thanksref{addr13}
\and
R.~Budnik\orcidlink{0000-0002-1963-9408}\thanksref{addr16}
\and
C.~Cai\thanksref{addr26}
\and
C.~Capelli\orcidlink{0000-0003-3330-621X}\thanksref{addr17}
\and
J.~M.~R.~Cardoso\orcidlink{0000-0002-8832-8208}\thanksref{addr2}
\and
A.~P.~Cimental~Ch\'avez\orcidlink{0009-0004-9605-5985}\thanksref{addr17}
\and
A.~P.~Colijn\orcidlink{0000-0002-3118-5197}\thanksref{addr8}
\and
J.~Conrad\orcidlink{0000-0001-9984-4411}\thanksref{addr12}
\and
J.~J.~Cuenca-Garc\'ia\orcidlink{0000-0002-3869-7398}\thanksref{addr17}
% ...
(XENON Collaboration\thanksref{xenonemail}). }
\newcommand{\bologna}{Department of Physics and Astronomy, University of Bologna and INFN-Bologna, 40126 Bologna, Italy}
\newcommand{\chicago}{Department of Physics, Enrico Fermi Institute \& Kavli Institute for Cosmological Physics, University of Chicago, Chicago, IL 60637, USA}
\newcommand{\coimbra}{LIBPhys, Department of Physics, University of Coimbra, 3004-516 Coimbra, Portugal}
\newcommand{\columbia}{Physics Department, Columbia University, New York, NY 10027, USA}
\newcommand{\lngs}{INFN-Laboratori Nazionali del Gran Sasso and Gran Sasso Science Institute, 67100 L'Aquila, Italy}
\newcommand{\mainz}{Institut f\"ur Physik \& Exzellenzcluster PRISMA$^{+}$, Johannes Gutenberg-Universit\"at Mainz, 55099 Mainz, Germany}
\newcommand{\mpik}{Max-Planck-Institut f\"ur Kernphysik, 69117 Heidelberg, Germany}
\newcommand{\munster}{Institut f\"ur Kernphysik, University of M\"unster, 48149 M\"unster, Germany}
\newcommand{\nikhef}{Nikhef and the University of Amsterdam, Science Park, 1098XG Amsterdam, Netherlands}
\newcommand{\nyuad}{New York University Abu Dhabi - Center for Astro, Particle and Planetary Physics, Abu Dhabi, United Arab Emirates}
\newcommand{\purdue}{Department of Physics and Astronomy, Purdue University, West Lafayette, IN 47907, USA}
\newcommand{\rice}{Department of Physics and Astronomy, Rice University, Houston, TX 77005, USA}
\newcommand{\stockholm}{Oskar Klein Centre, Department of Physics, Stockholm University, AlbaNova, Stockholm SE-10691, Sweden}
\newcommand{\subatech}{SUBATECH, IMT Atlantique, CNRS/IN2P3, Nantes Universit\'e, Nantes 44307, France}
\newcommand{\torino}{INAF-Astrophysical Observatory of Torino, Department of Physics, University  of  Torino and  INFN-Torino,  10125  Torino,  Italy}
\newcommand{\ucsd}{Department of Physics, University of California San Diego, La Jolla, CA 92093, USA}
\newcommand{\wis}{Department of Particle Physics and Astrophysics, Weizmann Institute of Science, Rehovot 7610001, Israel}
\newcommand{\zurich}{Physik-Institut, University of Z\"urich, 8057  Z\"urich, Switzerland}
\newcommand{\paris}{LPNHE, Sorbonne Universit\'{e}, CNRS/IN2P3, 75005 Paris, France}
\newcommand{\freiburg}{Physikalisches Institut, Universit\"at Freiburg, 79104 Freiburg, Germany}
\newcommand{\napels}{Department of Physics ``Ettore Pancini'', University of Napoli and INFN-Napoli, 80126 Napoli, Italy}
\newcommand{\nagoya}{Kobayashi-Maskawa Institute for the Origin of Particles and the Universe, and Institute for Space-Earth Environmental Research, Nagoya University, Furo-cho, Chikusa-ku, Nagoya, Aichi 464-8602, Japan}
\newcommand{\laquila}{Department of Physics and Chemistry, University of L'Aquila, 67100 L'Aquila, Italy}
\newcommand{\tokyo}{Kamioka Observatory, Institute for Cosmic Ray Research, and Kavli Institute for the Physics and Mathematics of the Universe (WPI), University of Tokyo, Higashi-Mozumi, Kamioka, Hida, Gifu 506-1205, Japan}
\newcommand{\kobe}{Department of Physics, Kobe University, Kobe, Hyogo 657-8501, Japan}
\newcommand{\kit}{Institute for Astroparticle Physics, Karlsruhe Institute of Technology, 76021 Karlsruhe, Germany}
\newcommand{\tsinghua}{Department of Physics \& Center for High Energy Physics, Tsinghua University, Beijing 100084, P.R. China}
\newcommand{\ferrara}{INFN-Ferrara and Dip. di Fisica e Scienze della Terra, Universit\`a di Ferrara, 44122 Ferrara, Italy}
\newcommand{\groningen}{Nikhef and the University of Groningen, Van Swinderen Institute, 9747AG Groningen, Netherlands}
\newcommand{\westlake}{Department of Physics, School of Science, Westlake University, Hangzhou 310030, P.R. China}
\newcommand{\shenzhen}{School of Science and Engineering, The Chinese University of Hong Kong, Shenzhen, Guangdong, 518172, P.R. China}
\newcommand{\coimbrapoli}{Coimbra Polytechnic - ISEC, 3030-199 Coimbra, Portugal}
\newcommand{\uniheidelberg}{Physikalisches Institut, Universit\"at Heidelberg, Heidelberg, Germany}
\newcommand{\roma}{INFN-Roma Tre, 00146 Roma, Italy}
\newcommand{\bucknell}{Department of Physics \& Astronomy, Bucknell University, Lewisburg, PA, USA}
\authorrunning{XENON Collaboration}
\thankstext{addr33}{Also at \roma}
\thankstext{addr34}{Now at \bucknell}
\thankstext{addr31}{Also at \coimbrapoli}
\thankstext{addr35}{Now at \munster}
\thankstext{xenonemail}{\texttt{xenon@lngs.infn.it}}
\thankstext{email0}{\texttt{andrea.mancuso@bo.infn.it}}
\thankstext{email1}{\texttt{dwenz@uni-muenster.de}}
\institute{\hypertarget{addr3}{\columbia}\label{addr3}
\and
\hypertarget{addr28}{\groningen}\label{addr28}
\and
\hypertarget{addr23}{\tokyo}\label{addr23}
\and
\hypertarget{addr18}{\paris}\label{addr18}
\and
\hypertarget{addr7}{\munster}\label{addr7}
\and
\hypertarget{addr14}{\torino}\label{addr14}
\and
\hypertarget{addr4}{\lngs}\label{addr4}
\and
\hypertarget{addr1}{\chicago}\label{addr1}
\and
\hypertarget{addr9}{\nyuad}\label{addr9}
\and
\hypertarget{addr17}{\zurich}\label{addr17}
\and
\hypertarget{addr13}{\subatech}\label{addr13}
\and
\hypertarget{addr0}{\bologna}\label{addr0}
\and
\hypertarget{addr6}{\mpik}\label{addr6}
\and
\hypertarget{addr19}{\freiburg}\label{addr19}
\and
\hypertarget{addr16}{\wis}\label{addr16}
\and
\hypertarget{addr26}{\tsinghua}\label{addr26}
\and
\hypertarget{addr2}{\coimbra}\label{addr2}
\and
\hypertarget{addr8}{\nikhef}\label{addr8}
\and
\hypertarget{addr12}{\stockholm}\label{addr12}
\and
\hypertarget{addr5}{\mainz}\label{addr5}
\and
\hypertarget{addr22}{\laquila}\label{addr22}
\and
\hypertarget{addr25}{\kit}\label{addr25}
\and
\hypertarget{addr10}{\purdue}\label{addr10}
\and
\hypertarget{addr20}{\napels}\label{addr20}
\and
\hypertarget{addr11}{\rice}\label{addr11}
\and
\hypertarget{addr15}{\ucsd}\label{addr15}
\and
\hypertarget{addr21}{\nagoya}\label{addr21}
\and
\hypertarget{addr29}{\westlake}\label{addr29}
\and
\hypertarget{addr24}{\kobe}\label{addr24}
\and
\hypertarget{addr27}{\ferrara}\label{addr27}
\and
\hypertarget{addr30}{\shenzhen}\label{addr30}
}

\date{Received: date / Accepted: date}

\onecolumn 
\maketitle
\twocolumn


\begin{abstract}%
Radiogenic neutrons emitted by detector materials are one of the most challenging backgrounds for the direct search of dark matter in the form of weakly interacting massive particles (WIMPs). 
To mitigate this background, the XENONnT experiment is equipped with a novel gadolinium-doped water Cherenkov detector, which encloses the xenon dual-phase time projection chamber (TPC).
The neutron veto (NV) tags neutrons via their capture on gadolinium or hydrogen, which release $\gamma$-rays that are subsequently detected as Cherenkov light.
In this work, we present the key features and the first results of the XENONnT NV when operated with demineralized water in the initial phase of the experiment. 
Its efficiency for detecting neutrons is \qty{82\pm1}{\percent}, the highest neutron detection efficiency achieved in a water Cherenkov detector. 
This enables a high efficiency of \qty{53\pm3}{\percent} for the tagging of WIMP-like neutron signals, inside a tagging time window of \qty{250}{\mus} between TPC and NV, leading to a livetime loss of \qty{1.6}{\percent} during the first science run of XENONnT.

\end{abstract}

\section{Introduction}
\label{intro}

The XENON project aims at the direct detection of dark matter \cite{direct_detection}, primarily in the form of weakly interacting massive particles (WIMPs) \cite{wimps}. 
The project consists of a series of dual-phase liquid-gas xenon time projection chambers (TPC) with increasing active mass and decreasing background, operated underground at the INFN Laboratori Nazionali del Gran Sasso, Italy \cite{XENON:2010xwm,XENON100:2011cza,XENON:2017lvq}.
The current experiment, XENONnT \cite{XENON:2024InstrumentPaper}, is installed at LNGS since 2020 as an upgrade of its predecessor XENON1T \cite{XENON:2017lvq,XENON:2018voc}, with commissioning and a first science run (SR0) completed during 2021 \cite{XENON:2022wimp,XENON:2022lower}. 
XENONnT has an active liquid xenon (LXe) target mass of \qty{5.9}{t}.

Particles interacting inside the LXe target produce scintillation photons and free electrons. 
The scintillation photons are detected as a first signal (S1) by two arrays of photomultiplier tubes (PMTs) located at the top and bottom of the TPC. 
A set of wire electrodes allows to drift and extract the free electrons from the liquid into the gaseous xenon phase, where the larger extraction field accelerates them to produce an electroluminescence signal, called S2 \cite{XENON:2024InstrumentPaper}.

Two principal categories of signals are observed in LXe TPCs: electronic recoil signals (ER) mostly originating from $\beta$-electrons and $\gamma$-rays interacting with the shell electrons of xenon, and nuclear (NR) recoils induced by coherent elastic scattering of neutral particles, like WIMPs, neutrons or neutrinos, with the xenon nuclei  \cite{XENON:2024xgd}. 
Thus, NR background signals induced by radiogenic neutrons originating from spontaneous fission and alpha-n reactions \cite{Cano-Ott:2024elo}, or cosmogenic neutrons induced by muons \cite{Pec:2023yic}, are generally a more problematic background since they are indistinguishable from WIMPs. 
For WIMP masses below \qty{5}{GeV \per c \squared}, coherent elastic neutrino-nucleus scattering of solar \isotope[8]{B} neutrinos are an additional irreducible NR background \cite{XENON:2024ijk,cevns_search,DARWIN:2016}. 
With the introduction of a water shield and active muon veto (MV) surrounding the TPC cryostat in XENON1T \cite{XENON:2017lvq,XENON1T:2014eqx}, the impact of NRs from cosmogenic neutrons on the WIMP sensitivity was reduced. 
Thanks to the unprecedentedly low ER background reached in XENONnT \cite{XENON:2022lower}, the NR background from radiogenic neutrons became more relevant in the search for WIMP dark matter, and thus a mitigation strategy is mandatory.

Most of the radiogenic neutrons that produce a signal inside the TPC exit the cryostat, where they are moderated mostly by protons of the surrounding water and eventually captured after a medium-dependent delay, emitting one or multiple $\gamma$-rays in the subsequent nuclear de-excitation. 
These $\gamma$-rays can be detected via Cherenkov photons emitted by fast electrons originating from Compton scattering.
Thus, enclosing the inner volume of the water tank with a highly light-sensitive detector, these Cherenkov signals can be used to effectively veto neutron-induced NR signals.
This technique of tagging neutrons was already successfully deployed by large-scale water-based neutrino experiments like Super-Kamiokande and SNO+ \cite{sk_neutron_capture_water,sno_ncapture}.

In this work, we describe the performance of the XENONnT neutron veto (NV) during its first de\-min\-er\-alized-water phase in which neutrons were tagged through their capture on hydrogen releasing a \qty{2.2}{MeV} $\gamma$-ray.
In section \ref{sec:detector_description} the design, the installation of the neutron veto, and its calibration tools are described.
Electronics and data acquisition are summarized in section \ref{sec:daq}, followed by the data processing pipeline in section \ref{sec:processing}.
Sections \ref{sec:pmt_calibration} discuss the performance of the PMTs, and detector stability.
The calibration of the NV efficiency to detect and tag neutrons is discussed in section \ref{sec:deteff}.  
Finally, section \ref{sec:nv_in_sr0} discusses the impact of the NV in the SR0 WIMP search before a summary and outlook are given. 

\section{The XENONnT neutron veto}
\label{sec:detector_description}
\subsection{Detector description}
\begin{figure*}[tb]
\centering 
\includegraphics[width=.95\textwidth]{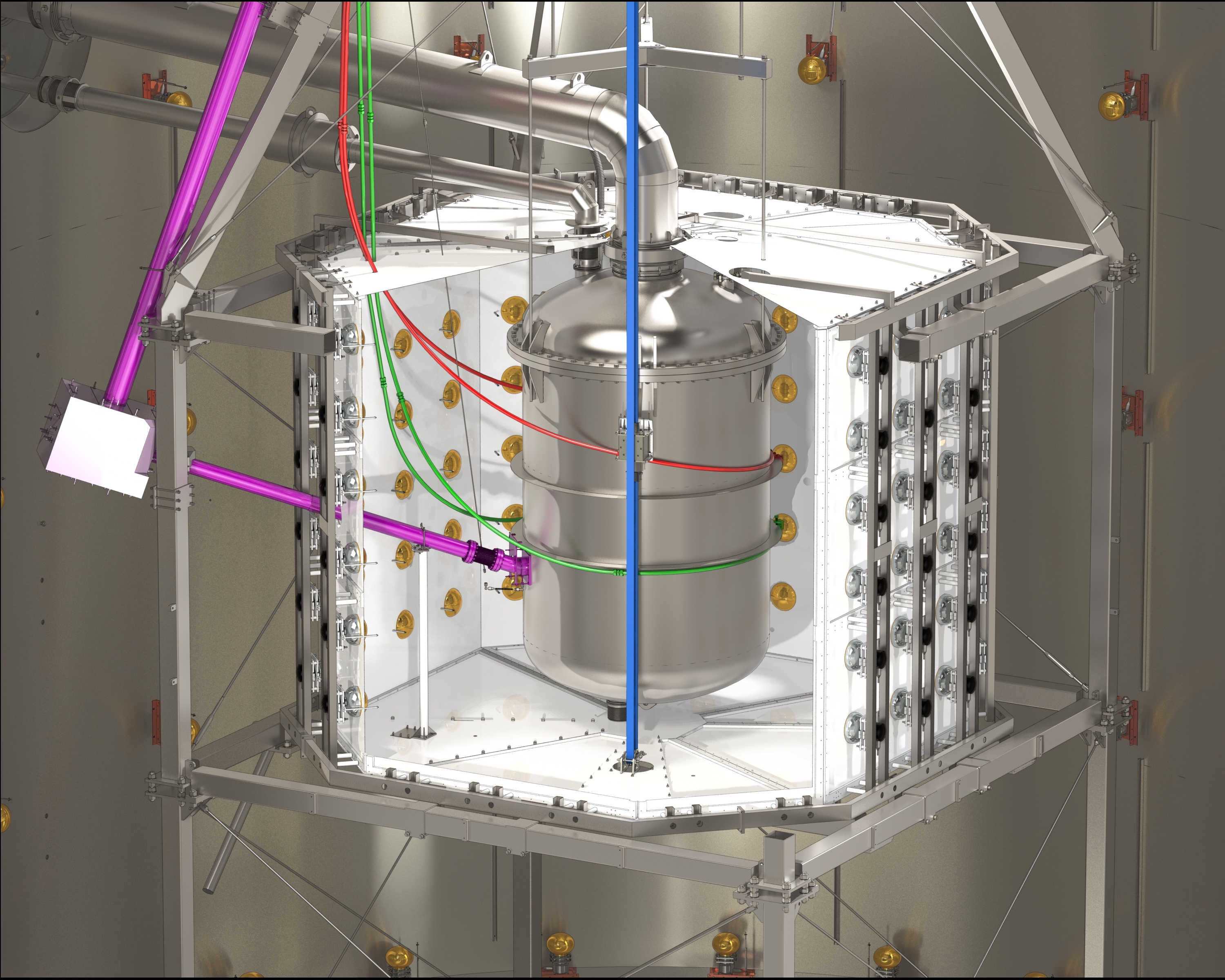}
\hfill
\caption{\label{fig:nveto-design}CAD rendering of the NV surrounding the TPC cryostat in the center of the XENONnT water tank. Its main elements are the support structure (grey), reflector panels (white), and the 120 PMTs (orange). The main components of the calibration system entering the NV are also shown: neutron generator pipe (purple), I-belt (blue), and U-tubes (red and green). The reflector foils attached to the cryostat were deliberately removed in this drawing.}
\label{fig:nv_cad_drawing}
\end{figure*}

The XENONnT NV is installed inside the \qty{700}{\cubic \meter} water tank of the muon veto, around the cryostat containing the TPC, as shown in Fig.\,\ref{fig:nveto-design}. It encloses a \qty{33}{\cubic \meter} volume, confined by high-reflectivity panels made of expanded polytetrafluoroethylene (ePTFE).
The panels form an octagonal prism with a height of \qty{3.16}{\meter}, and a side length of \qty{2.02}{\meter} and \qty{1.22}{\meter}. Wide and narrow sides are placed at a distance of \qty{1.87}{\meter} and \qty{2.04}{\meter} from the center, respectively.
The floor of the NV is made of eight triangular-shaped sections closing the whole surface, \qty{26}{\centi \meter} below the bottom of the TPC cryostat.
The ceiling is made of eight individual triangles, anchored on a ring around the main pipe at the center, \qty{20}{\centi \meter} above the cryostat dome. 
Each triangle is inclined by \qty{5}{\degree}, slightly decreasing towards the outer edge of the prism.
The ceiling features three ports: two of them are circular with a \qty{23}{\centi \meter} diameter, allowing the insertion of a neutron generator. The third port is rectangular ($23 \times 27$\,\SI{}{\square \centi \meter}) which allows the insertion of a tungsten shield including a \isotope[88]{Y}{}Be photon-neutron source for the calibration of the TPC \cite{XENON:2024InstrumentPaper}.
These ports are the only areas where photons can travel from the NV into the MV volume or vice-versa.

The \qty{1.5}{\milli \meter} thick ePTFE foil was chosen after a careful comparison of the reflectivity of several materials like Tyvek (1073B, 1073D, 1082D), Lumirror, Spectralon (9833, 9838), PTFE, and ePTFE, from which ePTFE presented the best reflectivity of $>$\qty{99}{\percent} for wavelengths greater than \qty{300}{\nano \meter} \cite{ePTFE}. 
The surface of the cryostat is covered with the same ePTFE foils, fixed with stainless steel studs and strings.
For the NV walls, the ePTFE foils are stretched and held in place with plastic frames, which are anchored on the cryostat support structure by thin stainless steel bars, to minimize backgrounds from NV-related structural materials.
Like all other parts of the experiment, the components of NV  underwent low-background screening to assess and validate their radiopurity levels \cite{XENON:2021mrg}.

Cherenkov photons produced by charged particles in the NV are detected by 120 high quantum efficiency (with a maximum of \qty{39}{\percent} at \qty{350}{\nano \meter}, on average), low-radioactivity, 8" PMTs, Hamamatsu R5912-100-10 WA-D30-SEL-Assy, sealed to be operated in water.
They are arranged in 20 columns of 6 PMTs each, vertically spaced by about \qty{45}{\centi \meter}.
Wide sides host three PMT columns, while narrow sides host two. 
The PMTs are mounted on the same steel bars holding the panels, with only the glass window protruding inside the NV volume. In this way, the impact of backgrounds produced by the radioactivity of the PMT body is minimized inside the NV.

\subsection{Light calibration tools}
\label{sec:lightcal}
\label{sec:diffuser_balls_and_ref_mon}
The NV is equipped with a set of optical calibration tools to continuously monitor the performance of the detector and its PMT response.
Among these tools are optical fibers leading to each PMT, four ``diffuser balls'' mounted around the TPC cryostat, and a "reflectivity monitor" set up to illuminate the NV ePTFE walls.

A system of 120 optical fibers is implemented to characterize and monitor the single photoelectron (SPE) response of the NV PMTs. 
A PTFE diffuser at the end of each fiber provides the optical coupling to its respective PMT. 
Outside the water tank, each bundle of 6 fibers belonging to the same column of PMTs is illuminated by a blue (\qty{470}{\nano \meter}) LED. The 20 LEDs are powered by Quantum Composer 9530 fast pulse generators that send an external trigger signal to the data acquisition at each LED pulse. 

A system of four diffuser balls illuminated by short laser pulses characterizes the timing and optical properties of the NV, such as water transparency and ePTFE surface reflectivity. 
They are mounted on the cryostat at equal heights near the center of the NV and equal angular distances from each other, illuminating one-quarter of the NV volume each.
The diffuser balls follow the design based on the one developed for the XENON1T MV \cite{christopher_geis}, but optimized for more precise photon timing. 
Each diffuser ball was tested for its isotropic light response and photon timing. 
The time spread of the diffuser balls ranges between \qty{4}{\nano \second} and \qty{6}{\nano \second} in air. 
Each diffuser ball can be illuminated separately using a picosecond laser (PILAS DX, NKT Photonics) equipped with a \qty{448\pm3}{\nano \meter} laser diode (PIL1-044-40FC), and an optical switch. 
An optical attenuator allows the adjustment of the laser light intensity down to a level where individual PMTs detect a single to a few photons. 

An additional reflectivity monitor is installed at a\-bout half height on the inside of one of the NV's octagon sides to monitor the optical properties of the NV. 
The monitor consists of four optical quartz fibers (Thorlabs UM22-200, polyimide coated), protected in a stainless steel tube, and fed by another picosecond laser (LDB-200, Tama Electronics) that emit photons at a wavelength of \qty{375\pm3}{\nano \meter}. 
This wavelength was chosen to probe near the maximum of the product of the NV's PMTs quantum efficiency and Cherenkov emission spectrum in water.
An optical switch is used to select one of the four fibers, two of which point upwards to illuminate the ceiling of the NV, while the other two are oriented downwards pointing to its floor.

\section{Electronics, DAQ, and time synchronization }
\label{sec:daq} 
\begin{figure*}[tb]
\centering 
\includegraphics[width=0.8\textwidth]{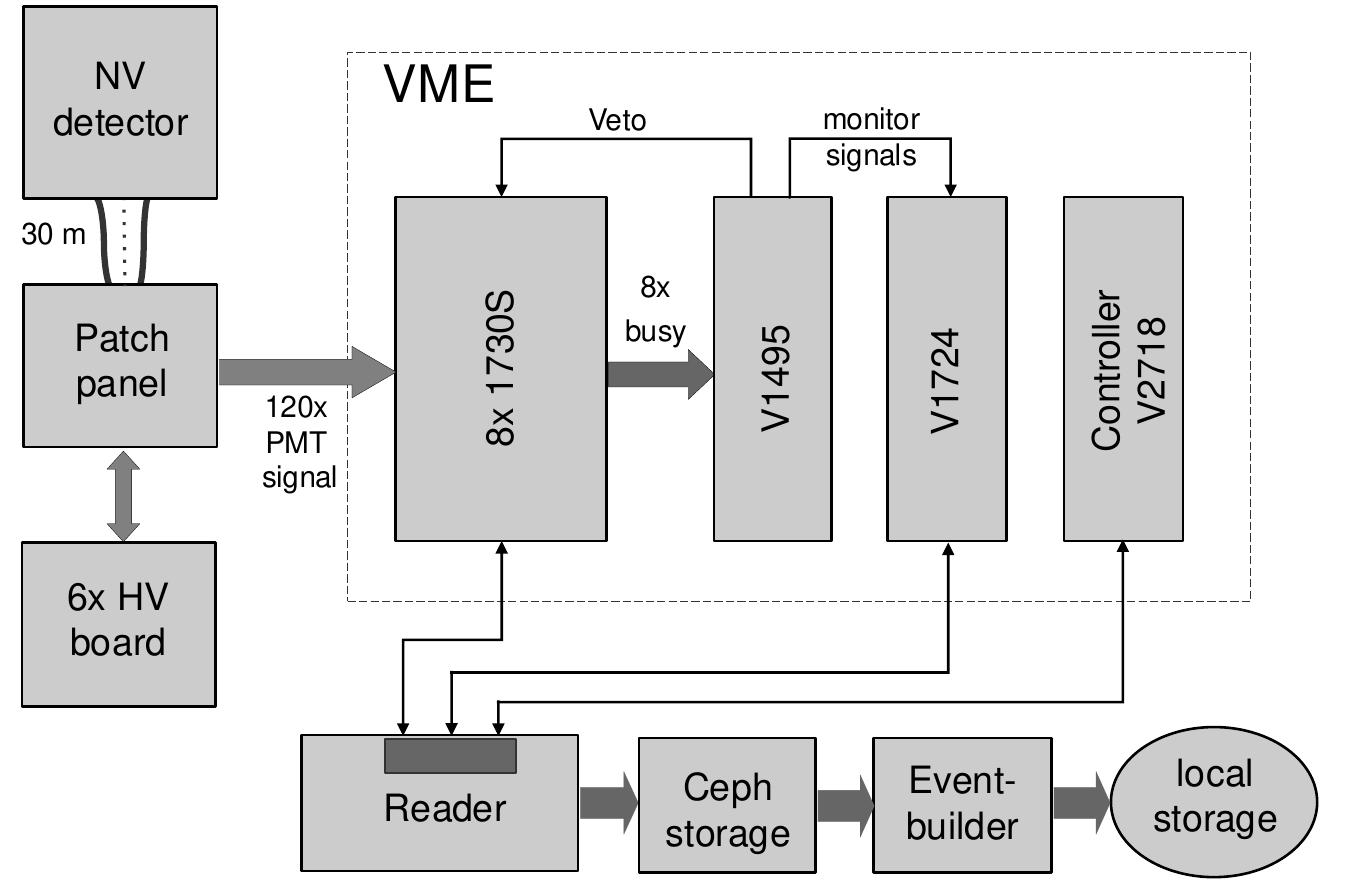}
\hfill
\caption{\label{fig:DAQscheme}NV DAQ scheme. PMT signals are digitized by the V1730S modules with a sampling rate of \qty{500}{MHz}. Data is read out from the digitizers by the reader server and written to a common (Ceph) storage disk available to the event-builder processing. The busy logic and the acquisition monitor functionalities are handled by V1495 and V1724 modules, respectively.}
\end{figure*}

The NV data acquisition is a fully integrated subsystem of the XENONnT DAQ system, described in \cite{XENON:2023daq}. 
It comprises its own set of electronics for digitization, data collection, and online processing.
The NV DAQ, as that of the TPC, is designed around a triggerless data collection scheme based on a readout system of independent channels, which allows the acquisition of each self-triggered PMT signal.
A schematic block diagram of the NV DAQ is shown in Fig. \ref{fig:DAQscheme}.
PMT coaxial cables for signal and high voltage (HV) are routed from the water tank to the rack of the NV electronics.
The PMT signals are connected to the front-end electronics employing a panel feedthrough. The positive PMT HV (\qty{+1.5}{\kilo \volt}, \qty{3.5}{\milli \ampere}) supplied by CAEN SY4527/A7435SP modules, is low-pass filtered to reduce high frequency noise ($>$ MHz).
Eight state-of-the-art waveform digitizer boards (CAEN V1730S) \cite{caen_general} acquire signals from the 120 PMTs. 
Each board houses 16 input channels, each containing a 14-bit \qty{500}{MHz} flash ADC. 
This sampling rate is five times faster than in the TPC, to reconstruct the fast arrival time of the directly impinging Cherenkov photons before diffuse reflection of the ePTFE walls dominates.  
The input section of the digitizers is adapted to \qty{50}{\Omega} and the dynamic range is set to \qty{2}{\volt}.

The V1730S digitizers are operated with the Digital Pulse Processing for Dynamic Acquisition Window (DPP-DAW) firmware, jointly developed by CAEN and XENON members for the readout of the XENON1T TPC \cite{XENON:2019bth}. 
Each channel handles the data acquisition in a self-trigger mode that allows storing the PMT waveform when the corresponding amplitude exceeds a programmed threshold with respect to the baseline, which itself is continuously computed. 
The waveform recording proceeds until the signal goes below the threshold with a dynamically enlarged window, up to a maximum of about \qty{1}{\micro s}. Most of the PMT channels have a self-trigger threshold set to \qty{15}{ADC} counts (ADCc) corresponding to an amplitude of about \qty{1.8}{\milli \volt} (about \qty{0.25}{\PE} for a PMT gain of $10^7$). 
An increased threshold of \qty{20}{ADC}c is used in ten channels with higher noise levels.

Each single pulse data frame, which includes a 48-bit trigger timestamp, a baseline, and all the waveform samples with variable length, is written to a large channel memory buffer designed as a First In, First Out system (FIFO). 
Pulses coming from different channels are recorded in different buffer FIFOs. Four optical links, with up to \qty{90}{MB \per \second} capacity each, connect the FIFOs to the CAEN A3818 PCIe card hosted in the ``Reader'' server shown in Fig. \ref{fig:DAQscheme}.
When the memory is completely full, the system triggers a ``busy'' signal which halts the data acquisition until some space has been freed up. 
When one or more boards are busy, an auxiliary board (CAEN V1495 in Fig. \ref{fig:DAQscheme}) inhibits the data acquisition for all boards by providing a veto signal. This board also produces the veto start and stop signals used by data acquisition monitor.
During SR0 WIMP-search data taking the NV DAQ has not exhibited any veto signal, thus inducing no dead time.

In science runs (WIMP-search, gamma, and neutron calibrations), the NV digitizers are operated in self-trigger mode. 
A controller board (CAEN V2718), that acts as a VME bridge, generates and distributes the start-of-acquisition to all the digitizers via logic fan-in/fan out modules to keep the timing synchronization within a few ns.
For optical calibrations, an external trigger signal is provided using the same logic.
An additional digitizer (CAEN V1724, with a sampling rate of \qty{100}{MHz}, the same used for the TPC and MV systems) records all these NV acquisition monitor signals (i.e. V1730 busies, veto start, and veto stop) to identify potential data loss due to the veto condition. 
To keep the V1730S board chain well synchronized, all digitizers share a common \qty{50}{MHz} clock signal, which is distributed to the V1724 module of the NV crate. 
An embedded Phase-Locked Loop outputs a \qty{62.5}{MHz} clock reference that is propagated through the V1730S board chain via {\it clk-in} and {\it clk-out} connectors, keeping the NV boards synchronized to within \qty{1}{\nano \second}. 

Custom software, originally developed for the TPC, was adapted for the NV V1730S boards.
It efficiently reads data in block transfers via the CAENVMElib, and writes it to a shared storage device accessible for online processing. 
During the NV DAQ commissioning, different run modes were validated, sustaining a data transfer rate of \qty{90}{MB \per \second} through one optical link.
In the final setup used in SR0, the data rates were about \qty{14}{MB \per \second} and \qty{40}{MB \per \second} in the self-trigger and calibration modes, respectively.

\section{Analysis software and data reconstruction}
\label{sec:processing}
The raw waveforms recorded by the digitizers are processed using the open-source software packages strax \cite{strax} and straxen \cite{straxen}, developed for the XENONnT experiment. 
Strax provides a general infrastructure to set up a processing chain for time-sorted peak-like data, while straxen contains XENONnT-specific code. 
The key components of this processing framework are so-called plugins, which divide the processing chain into many small logic steps. 

\begin{figure}[t]
\centering
\includegraphics[width=0.45\textwidth]{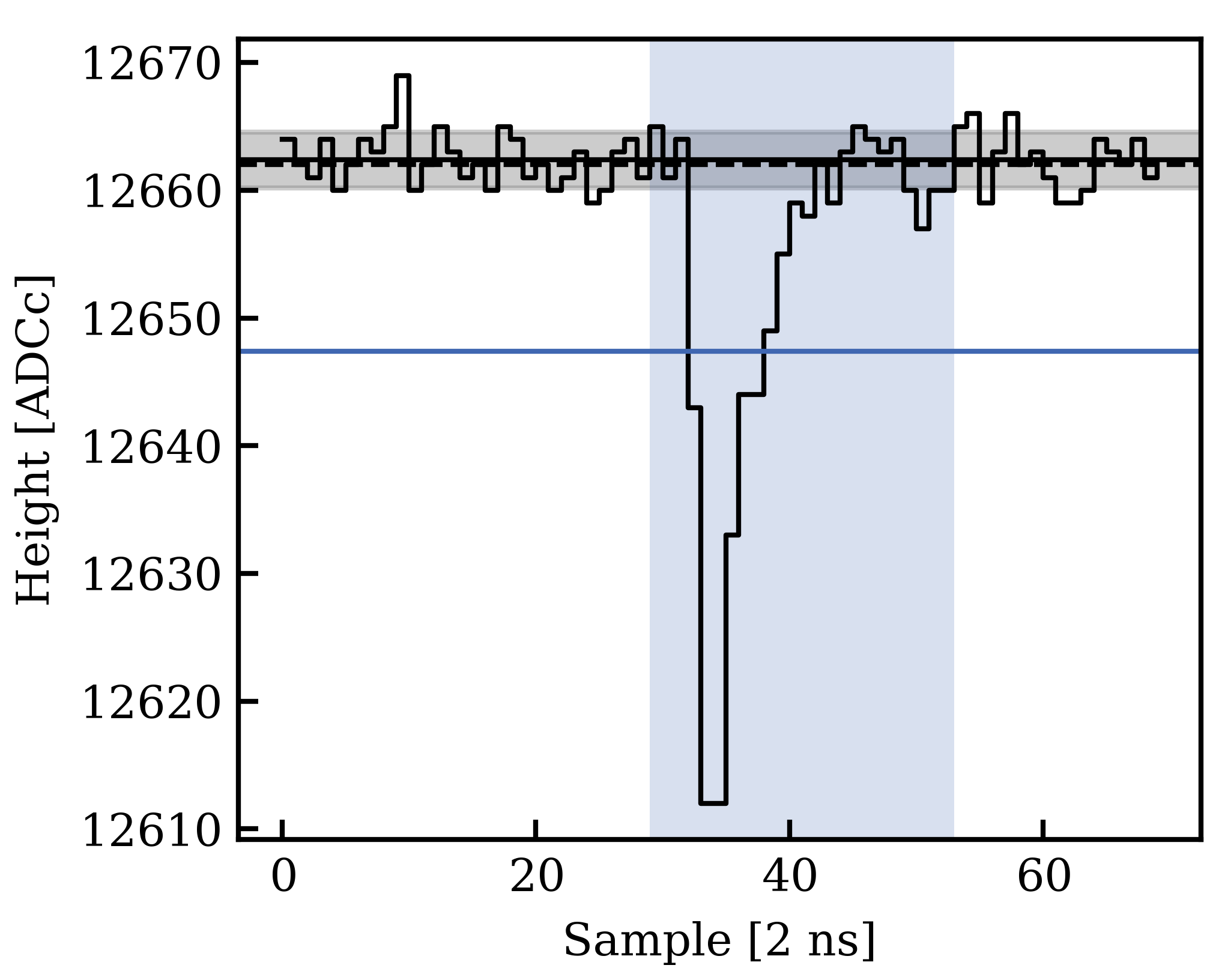}
\caption{Waveform of a single photoelectron signal recorded in PMT 2087. 
The black solid and dashed lines represent the baseline of the waveform as estimated by the digitizer and the processing software, respectively.
The black-shaded region indicates the baseline RMS. The blue horizontal line shows the hitfinder threshold, set to \qty{15}{ADC}c above baseline. 
All consecutive samples below this line are marked as a hit. 
The blue shaded region indicates the hit including the left and right extension, named ``hitlet'' in the straxen framework, as explained in the text.}
\label{fig_single_pmt_hit}  
\end{figure}

The processing chain of the NV data starts with the identification of the so-called hits for each of the recorded raw pulses. 
An example of a single PMT pulse is shown in Fig. \ref{fig_single_pmt_hit}.
Each pulse is baseline-corrected and flipped by subtracting the average pulse height computed on the first 26 samples of a pulse.  
Afterwards, PMT signals are identified by a hitfinder algorithm which searches for PMT ``hits'' by selecting consecutive samples above a fixed PMT-dependent threshold. 
The same thresholds as in the DAQ are used, leading to a typical hit length of three to five samples. Subsequently, a software trigger plugin discriminates physical events due to simultaneous Cherenkov photon emission from uncorrelated PMT dark counts by applying a moving coincidence window which searches for groups of at least 3 hits in a window of \qty{600}{ns}.
Each triggering window is extended by an additional \qty{150}{ns} long pre-trigger window. 
All raw PMT pulses partially overlapping with a triggering window are kept, while the remaining pulses can be deleted, effectively reducing the amount of raw data. 
During SR0 the software trigger was applied, but all data was kept to allow for extensive testing of the processing software. 

To include the leading edge and the tails of each PMT hit, hits are extended by a fixed number of samples before (3 samples) and after (15 samples) the thresh\-old-crossing sample, forming a so-called ``hitlet'', indicated in Fig. \ref{fig_single_pmt_hit} as a blue shaded region. 
Overlapping hitlets recorded by the same PMT are first concatenated before they are split at local prominences. 
The valley-to-peak ratio of the local prominences has to be below \qty{25}{\percent}, and the minimal height, defined per channel, must be of about \qty{20}{ADCc} to avoid a splitting of PMT signal tails due to baseline noise.
Afterwards, the basic properties of each hitlet like the amplitude, area, and shape parameters are computed. 
\begin{figure}[t]
\centering
\includegraphics[width=0.46\textwidth]{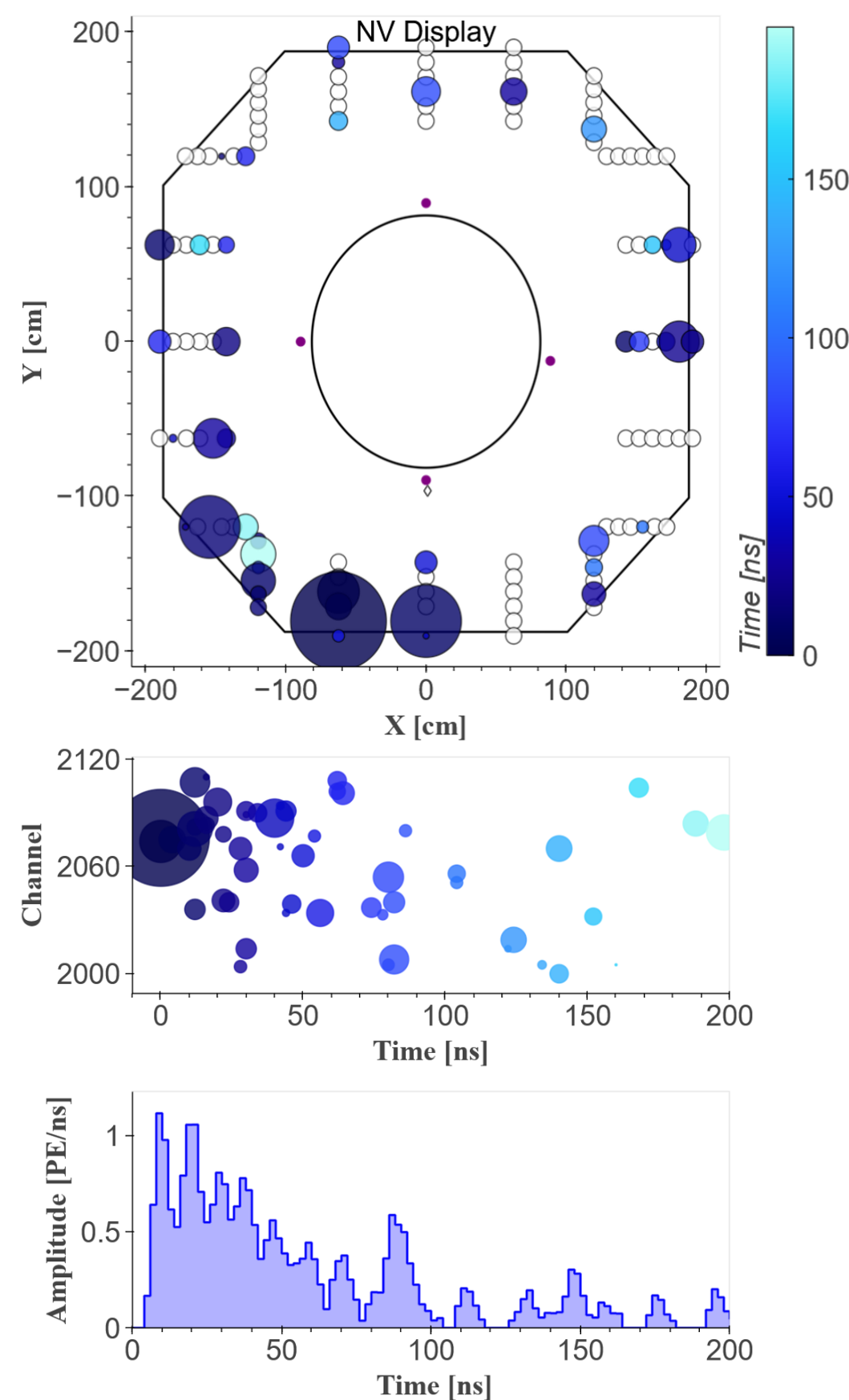}
\caption{Event display of a \qty{4.4}{MeV} $\gamma$-ray recorded during \isotope[241]{Am}Be calibration. Top: Two-dimensional projection of the NV. The outer wall of the cryostat (black circle), the four diffuser balls (purple dots) and the NV walls (black octagon) are shown. Each circle next to the octagon represents one of the NV PMTs with the innermost circle corresponding to the lowest PMT in a column. The size of the dot indicates the integrated charge detected by the PMT in the displayed event. The color encodes the arrival time of the first detected photon in each respective channel. Center: Arrival time of the individual hitlets for the given event, using the same legenda as in the top panel. Bottom: Summed waveform of the event in the NV. The event display in \cite{XENON:2024InstrumentPaper} shows for the same event the corresponding neutron interaction inside the TPC.}
\label{fig_event_display}  
\end{figure}

In the last step, hitlets are clustered into events using the same moving time-window coincidence also used in the software trigger, but with a tighter window of \qty{200}{ns},  motivated by the average arrival time spread of the Cherenkov signal of about \qty{60}{\nano \second}, discussed in detail in section \ref{sec:pmt_calibration}. 
An NV event is required to contain at least 3 hitlets. 
For each event, several properties are computed: the total signal area, the ``center time'', given by the area-weighted average arrival time of the constituent hitlets, and a simple position reconstruction. 
The center time was used as a reference for the optical properties of the NV, in addition to those monitored with the reflectivity and diffuser ball calibrations, using Cherenkov light directly.
The position of an event was estimated as the area-weighted average of all hitlets recorded within the first \qty{20}{\nano \second} of an event. 
Thus, it mostly includes photons that directly impinge on the PMTs before diffuse reflection on the ePTFE walls washes out any position information.
The PMT hit pattern and the time distribution of a typical NV signal are shown in Fig. \ref{fig_event_display}.

\section{PMT performance and detector stability}
\label{sec:pmt_calibration}
Relevant PMT parameters were monitored throughout commissioning and science data taking, such as stability of the baseline, single hit rate (dark rate), gains, and single photoelectron (SPE) acceptance.  

The baseline for each channel is calculated from the first \qty{26}{samples} of the waveforms acquired from the self-triggered data.
The mean and standard deviations of the baseline are determined. A typical value for the standard deviation is $\sigma_{\rm base}\approx \qty{2.5}{ADCc}$ resulting in an uncertainty in the baseline value of $\sigma_{\rm base}/\sqrt{26} \approx \qty{0.5}{ADCc}$, and motivating the hitfinder threshold of \qty{15}{ADCc}.

\begin{figure}
    \centering
    \includegraphics[width=0.49\textwidth]{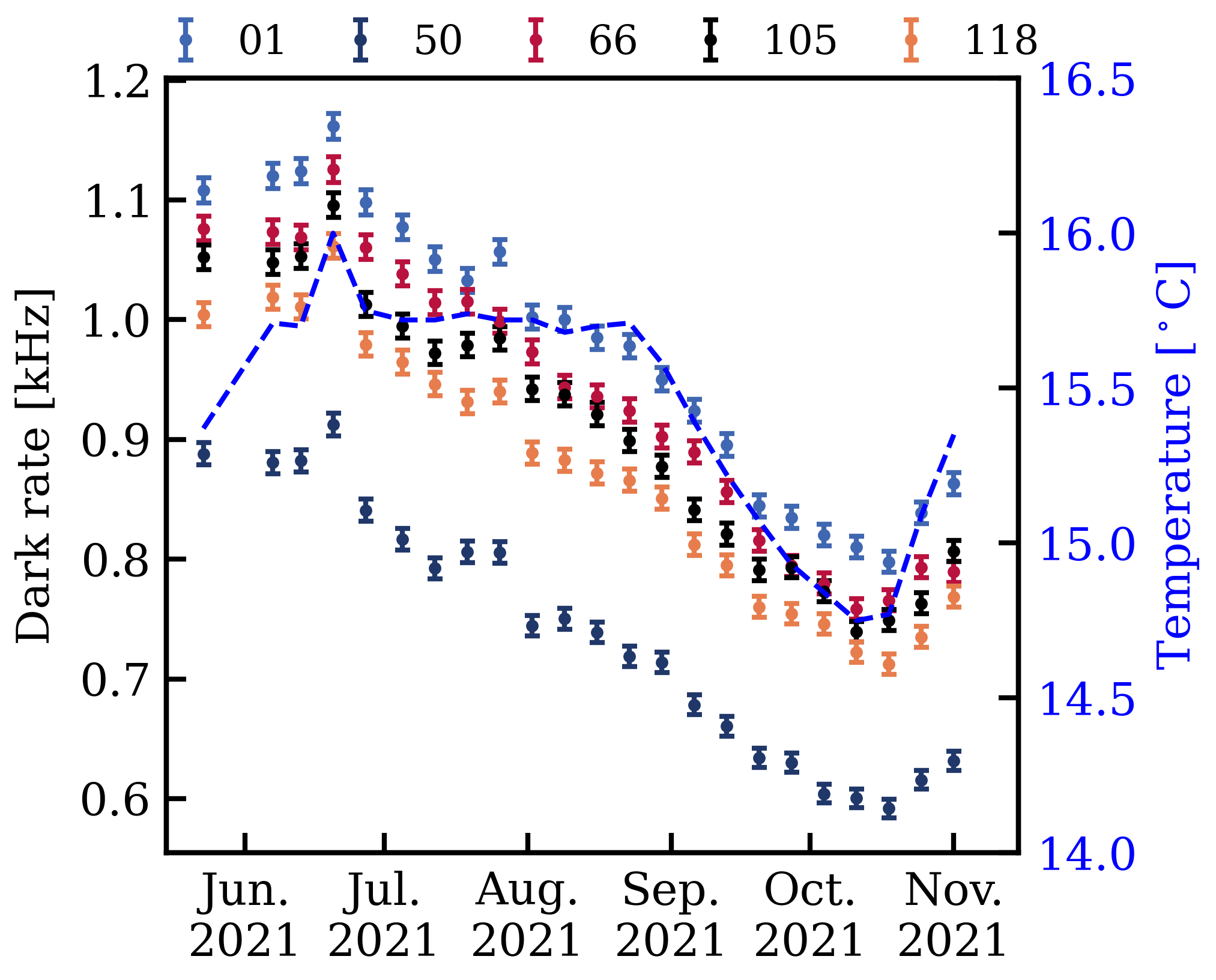}\\
    \caption{Evolution of the dark rate of six randomly chosen PMTs during SR0. The colors indicate different PMT channels. The temperature measured inside the demineralized water purification plant is shown as a blue-dashed line.}
    \label{fig:DR}
\end{figure}

The dark rate of a PMT was measured as the number of peaks over threshold in the self-triggered data in a given time interval: it ranges from about \qty{500}{Hz} to \qty{1500}{Hz}, with an average value of \qty{960\pm 4}{Hz} at a water temperature of $15.5^{\circ}$C. 
Fig. \ref{fig:DR} shows the sampled dark rate as a function of time during SR0 for a few randomly chosen PMTs. 
The dark rate is clearly correlated with the water temperature (blue dashed line), where the water follows the variation of the temperature of the underground experimental hall. 
The observation is compatible with the known temperature dependence of thermionic emission by bialkali photocathodes \cite{Hamamatsu:PMT:basics}, but had a negligible impact on the efficiency of the NV given the 5-fold PMT coincidence requirement for neutron tagging, discussed in section \ref{sec:deteff}.

\begin{figure}
    \centering
    \includegraphics[width=0.49\textwidth]{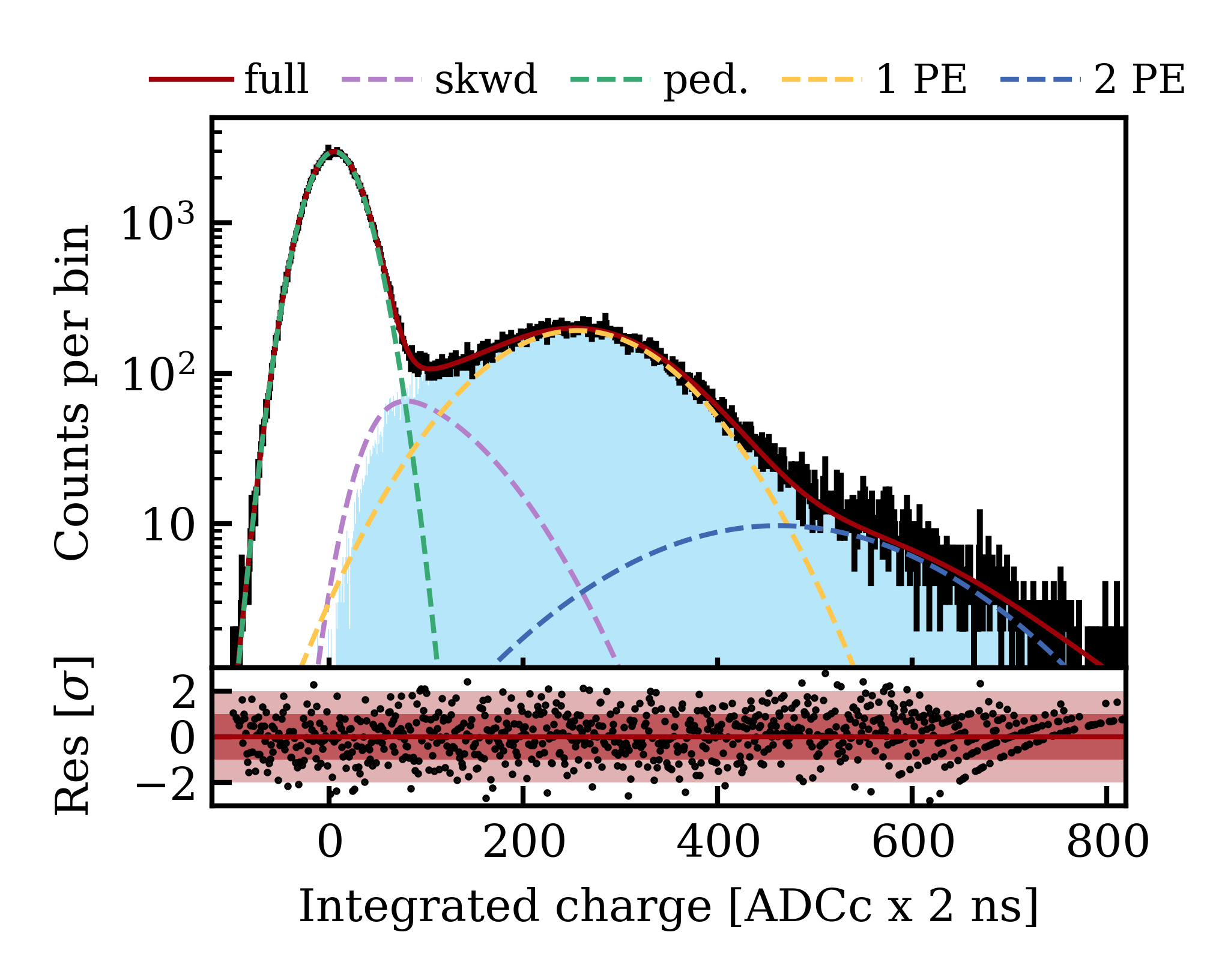}
    \caption{Charge spectrum for one PMT during the LED calibration. The fit function is superimposed, with contributions of the pedestal (ped.) in green, under amplified signal (skwd) in violet, single and double photoelectron (1 PE, 2 PE) in yellow and blue. The light blue histogram illustrates the signal distribution above 15 ADCc DAQ threshold. The bottom panel shows the residuals of the fit.}
    \label{fig:LED_spectrum}
\end{figure}

The gain of each PMT and its SPE acceptance are measured weekly using the light calibration tools described in section \ref{sec:diffuser_balls_and_ref_mon}, with an external trigger occurring at each LED illumination.   
Each acquired waveform was measured against the mean baseline value, obtained from a 30-sample window before the signal.
The charge of each waveform is obtained by identifying the sample $t_{\rm max}$ with the highest amplitude and integrating in a time interval [\qty{-20}{\nano \second}, \qty{+40}{\nano \second}] around $t_{\rm max}$. 
Fig. \ref{fig:LED_spectrum} shows a typical resulting charge spectrum of a single PMT. The pedestal, centered around zero and due to baseline-only events, is clearly visible. A valley follows it before a second, wider peak from fully amplified SPE signals. The tail at larger charge signals is due to double and multiple photoelectron events. 
We found that the pedestal peak is described more accurately by the sum of two Gaussian functions with equal sigma ($\sigma_{\rm  PED}$), but a different mean due to small baseline drifts.
The SPE component is made of two parts: fully amplified signals described by a Normal distribution, and underamplified signals modelled with a skew-Normal distribution. 
Underamplified signals are caused by electrons partially depositing their energy in the first dynode, collection inefficiency on the second and subsequent dynodes, or photons directly reaching the first dynode. 
Hence, the probability distribution function of the whole SPE component is:
\begin{equation}
\begin{array}{l l}
P_{\rm SPE}&(x) = f_{\rm FA} \cdot {\cal N}(x, \mu_{\rm FA},\sigma_{\rm FA}) \; \\
         &+(1-f_{\rm FA}) \cdot {\cal N}(x, \xi,\omega) \cdot \left[1+{\rm Erf}\left(\alpha \frac{(x-\xi)}{\sqrt{2}\omega}\right)\right]
\end{array}   
\end{equation}
where the first part represents the Normal distribution with parameters $\mu_{\rm FA}$ and $\sigma_{\rm FA}$ describing the fully amplified component which contributes a fraction $f_{\rm FA}$ of the total SPE distribution, and the second one is the skew-Normal distribution with location $\xi$, scale $\omega$ and shape parameter $\alpha$ modeling the partially amplified component.
The parameters $\mu_{\rm FA}$, $\sigma_{\rm FA}$, and $\xi$ are free parameters together with the fraction $f_{\rm FA}$ and the overall normalization, while $\omega$, and $\alpha$ are constrained in the fit by $\sigma_{\rm FA}$ and $\sigma_{\rm PED}$.
 
The multiple PE response is described by a set of Normal distributions with their means and standard deviations derived from the SPE response \cite{Saldanha:2016mkn}, therefore only the normalization of these distributions are free parameters.
During LED calibrations we kept the light emission low, minimizing the contribution of multi-PE pulses to the few percent level.

\begin{figure}
    \centering
    \includegraphics[width=0.49\textwidth]{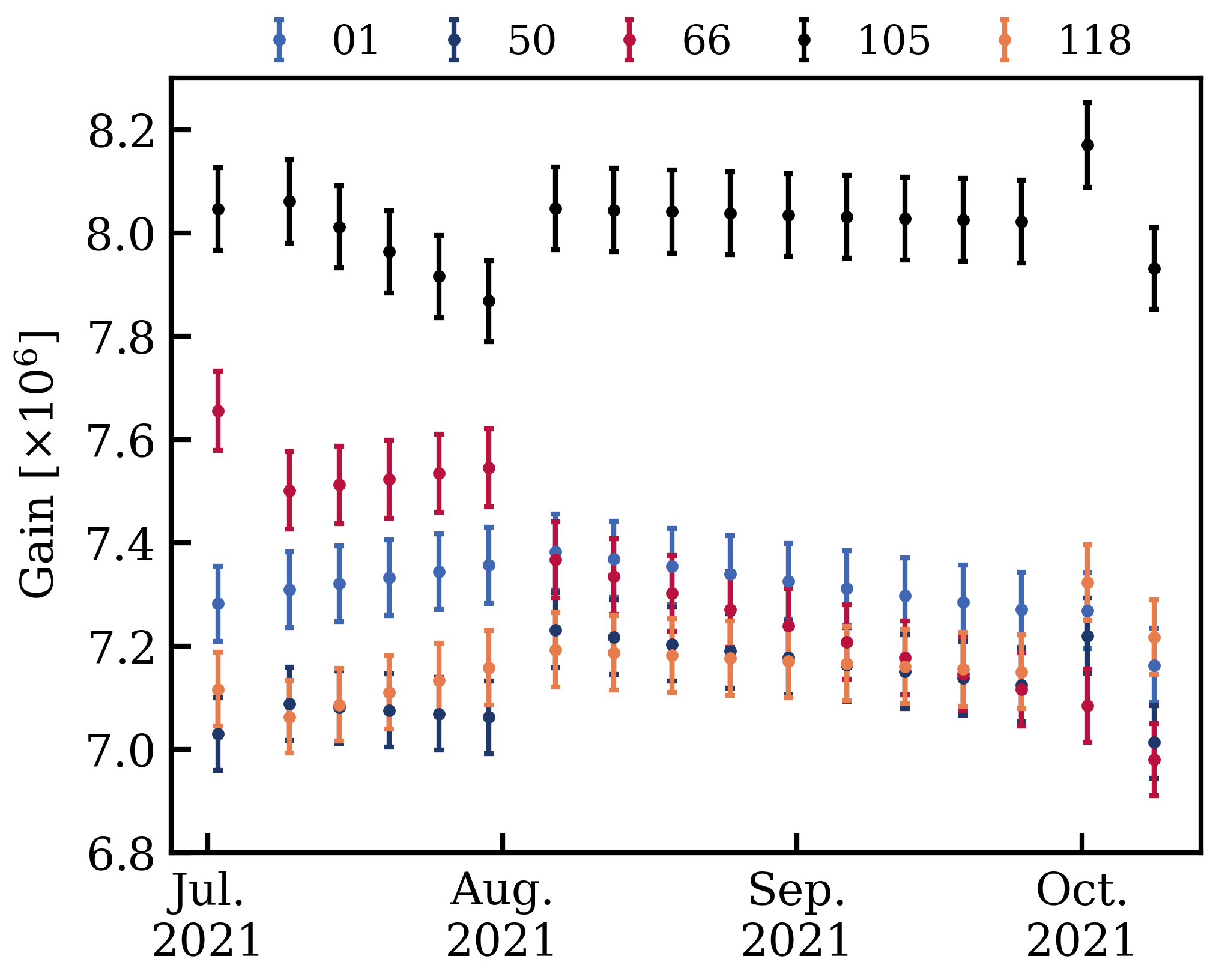}
    \caption{Evolution of the gain of six PMTs during SR0. Each point is the result of the weekly LED calibration performed during SR0.}
    \label{fig:PMTgain}
\end{figure}

From fits to these spectra, both PMT gain and SPE acceptance are determined. 
The gain is defined as the average value of the overall SPE distribution, which includes the partially amplified contribution. 
Fig.~\ref{fig:PMTgain} shows the gain values during SR0 for some channels (the same as shown also in Fig. \ref{fig:DR}). Typical gain values are about $7\times 10^{6}$ with a stability better than 5\%.

The SPE acceptance $\varepsilon$ is calculated as:
\begin{equation}
    \varepsilon(x, x_{th}) = \frac{ N(x_{th}) - \int [{\cal F}_{2\,\mathrm{PE}}(x)+ {\cal F}_{3\,\mathrm{PE}}(x)] dx}{ \int {\cal F}_{1\,\mathrm{PE}}(x) dx}\, ,
\end{equation}
where $N$ is the total number of events after applying the trigger threshold offline, and ${\cal F}_\mathrm{nPE}(x)$ represents the $n$-photoelectron contributions of the fit function. Averaging over all the 120 channels, the SPE acceptance has a mean value of 91\% and a standard deviation of 2.4\%. 

The evolution of the background rate in the NV is estimated via the event rate for different PMT coincidence requirements.
During the first months of detector commissioning an exponential decrease in this rate was observed, attributed to the decay of \isotope[222]{Rn} present in the water used to fill the tank, as reported in \cite{XENON:2024InstrumentPaper}. 

\begin{figure}
\centering
\includegraphics[width=0.49\textwidth]{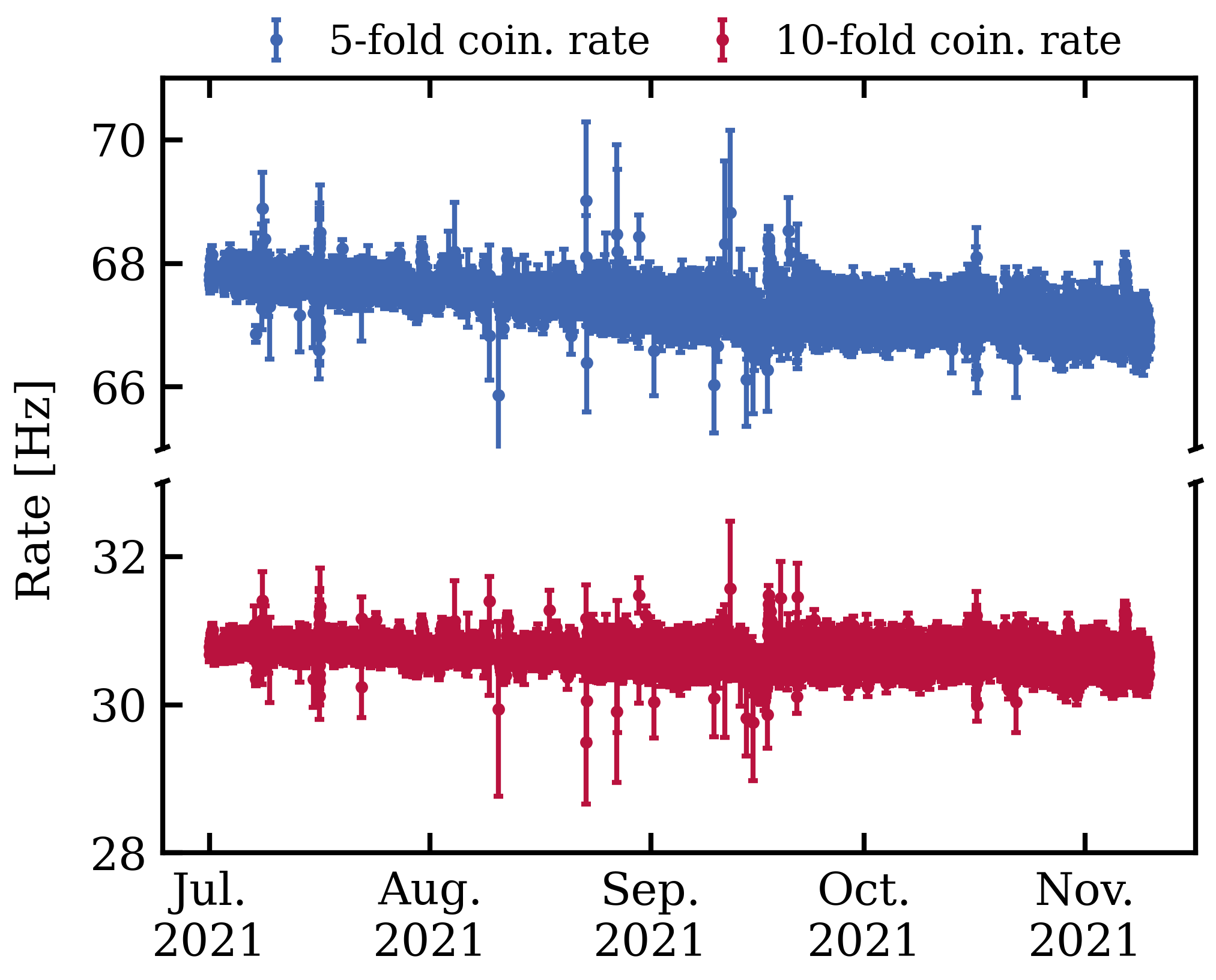}
\caption{Evolution of the 5- and 10-fold coincidence rate during SR0. Each point corresponds to a run. The 5-fold trend slightly decreases due to the PMT dark rate evolution.}
\label{coincidence_rate_sr0}  
\end{figure}

After commissioning, a slightly decreasing trend in the 5-fold coincidence rate was observed, as shown in Fig. \ref{coincidence_rate_sr0}. It can be associated with the overall decrease in the PMT dark rate caused by the change in water temperature. 
This trend is not observed in the 10-fold coincidence rate, suggesting that it originates from accidental coincidences and uncorrelated hits on the PMTs rather than being associated to physical events from radioactivity which are expected to be constant in time. 
Thus, all runs were accepted for SR0. 
The decreasing event rate is accounted for in the WIMP search by computing a time-dependent livetime loss due to the event rate in the NV. 

\begin{figure}
    \centering
    \includegraphics[width=0.99\linewidth]{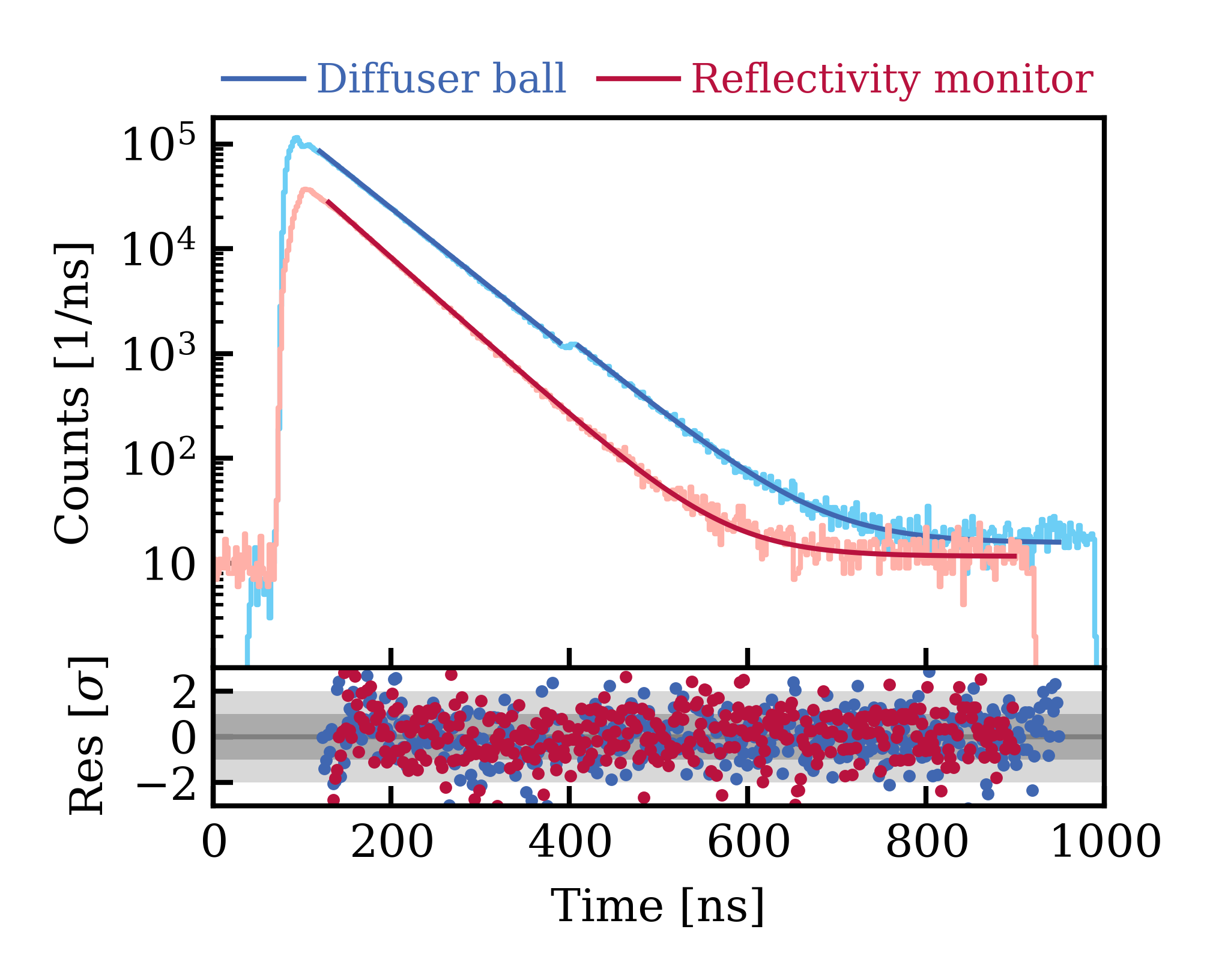}
    \caption{Arrival time of photons of all PMTs for an injected light signal using the reflectivity monitor (red) and a single diffuser ball (blue). The solid lines show the best-fit model for both distributions. The bottom panel shows the residuals of the fit.}
    \label{fig:diffuser_ball_time_spectrum}
\end{figure}

The optical properties of the NV were also monitored during SR0. The size and time distribution of the recorded Che\-ren\-kov signals not only depends on the PMT performance, but also on the total active photosensor area, water transparency, and wall reflectivity. 
The latter two are monitored by means of the systems described in section \ref{sec:lightcal}.
Fig. \ref{fig:diffuser_ball_time_spectrum} shows the arrival time distribution of photons, summed over all PMTs, for the reflectivity monitor and diffuser ball setups.
In the diffuser ball setup, total internal reflection within the glass fiber, with a length of \qty{30}{\meter}, leads to a second light emission after about \qty{300}{\nano \second}.   
The shape of the leading edge of both distributions depends slightly on the respective setup.  
Once photons are diffusely reflected by the ePTFE walls, both distributions follow a featureless exponential distribution with slightly different average-time parameter. 
Both distributions are chi-square-fitted using an exponential function on top of a constant background. 
Due to the internal light reflection, the fit of the diffuser ball data is divided into two regions. 
The average-time parameter of the optical photon signal for the diffuser balls (at $\lambda= \qty{448}{\nano \meter} $) is $\tau_{DB}=$\qty{63.9\pm0.5}{\nano \second}. The average time parameter for the reflectivity monitor setup (at $\lambda= \qty{375}{\nano \meter} $) is slightly shorter $\tau_{RM}=$\qty{57\pm1}{ns}, where the uncertainty is dominated by the systematic uncertainty obtained from the spread of the results in the four channels.

\begin{figure}
    \centering
    \includegraphics[width=0.49\textwidth]{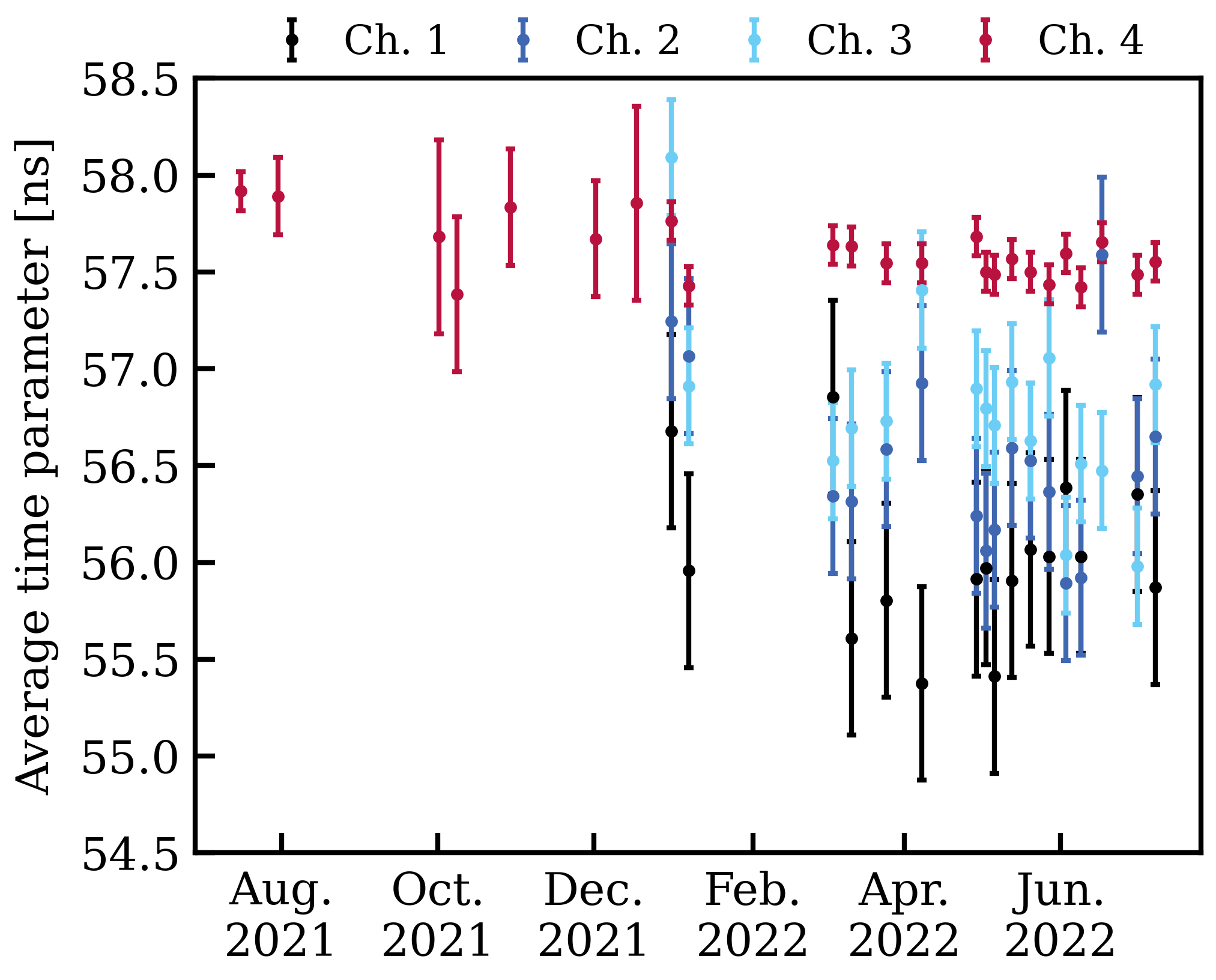}
    \caption{Evolution of the average time parameter in reflectivity monitor runs as a function of time, during SR0 and beyond.  During SR0, only Channel 4 was acquired. The spread between the four channels in 2022 was used to estimate the systematic uncertainty.}
    \label{fig:reflectivity_monitoring}
\end{figure}

The time stability of the reflectivity monitor measurements is shown in Fig. \ref{fig:reflectivity_monitoring}, where no significant change in $\tau_{RM}$ was found over one year of operations, demonstrating the stability of the light collection efficiency in the NV. 
 It should be noted that during SR0 only data with channel 4 of the reflectivity monitor was recorded for the whole run.
The average time parameter measured with the reflectivity monitor setup $\tau_{RM}$ is slightly lower than for the diffuser balls $\tau_{DB}$, which is interpreted as being due to different water transparency at the two laser's wavelengths. 
The ``center time'' of Cherenkov signals in the WIMP-search and calibration data suggests a time parameter consistent with the average value of the reflectivity monitor.

\section{Neutron veto efficiency}
\label{sec:deteff}
The key characteristic of the NV is its ability to detect and tag neutron signals. 
The detection efficiency determines the efficiency of the NV to detect an emitted neutron, while the tagging efficiency describes more specifically how efficiently the NV tags those neutrons that perform a single-scatter nuclear recoil (SSNR) in the region of interest for the WIMP search. Both efficiencies were measured using neutrons from an \isotope[241]{Am}{}\isotope[9]{}{Be} source (hereafter referred to as \isotope[241]{Am}Be), also used to calibrate the NR response of the TPC \cite{XENON:2022wimp}. 

\subsection{Calibration setup}
\label{ss:NVeff:setup}
During SR0, the TPC and NV were calibrated via external \isotope[232]{Th}- and \isotope[241]{Am}Be-calibration sources placed at different positions around the TPC cryostat, utilizing the U-tubes source delivery system shown in Fig. \ref{fig:nv_cad_drawing} and discussed in \cite{XENON:2024InstrumentPaper}. 
The \isotope[232]{Th}-source was used to test the calibration and source deployment procedure by mapping the spatial NV response of the emitted $\gamma$-rays when moving the source.
 
\begin{figure}
    \centering
    \includegraphics[width=0.49\textwidth]{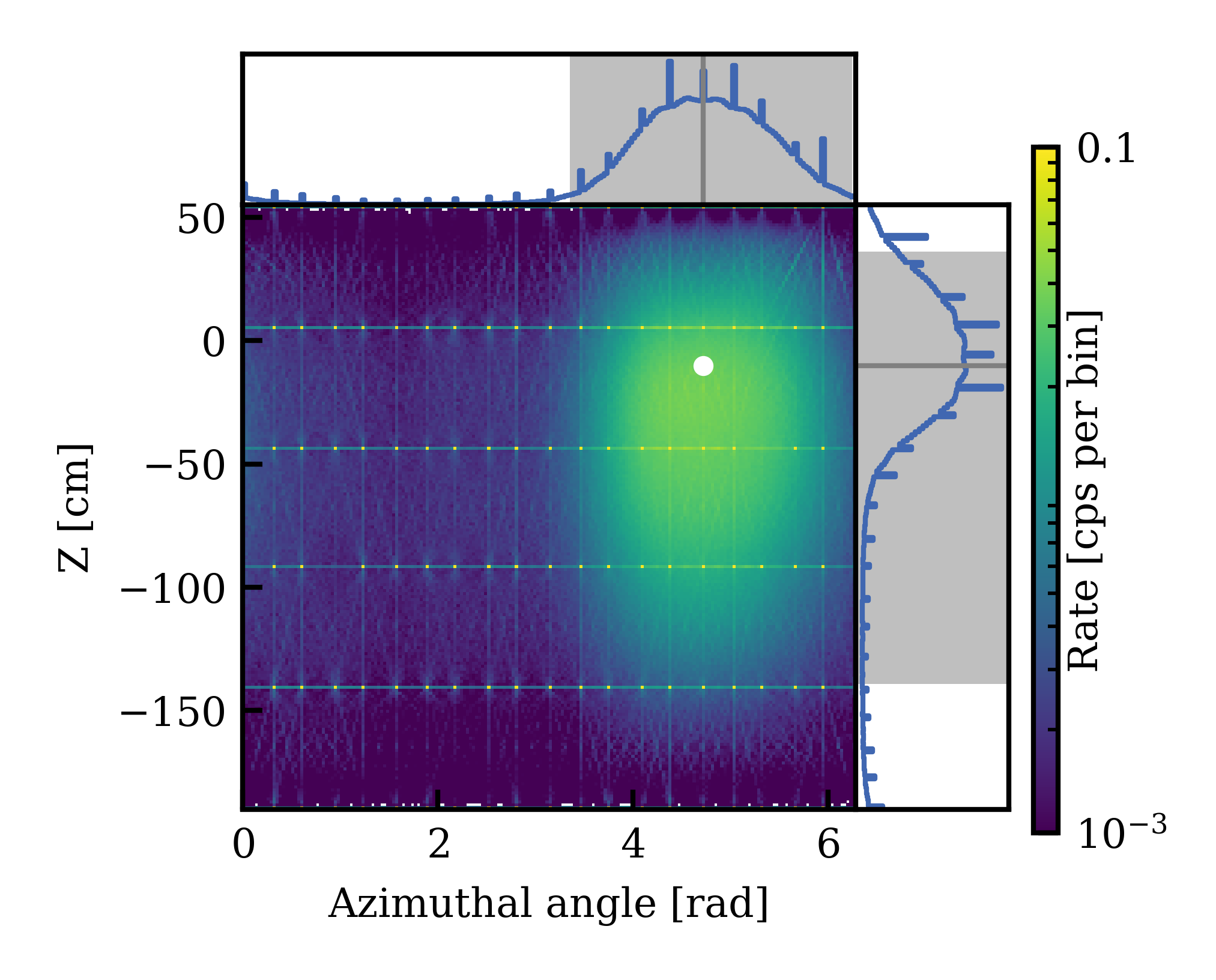}
    \caption{Spatial distribution of the reconstructed NV events from an \isotope[241]{Am}Be source located at the position indicated with the white dot. The blue curve shows the projections of the spatial distribution on the respective axes after a bin-wise background subtraction. The gray-shaded region indicates the central \qty{95}{\percent} region for the background corrected projections. The noticeable regular pattern of dots and lines is an artifact of the position reconstruction algorithm, where the dots indicate the positions of the PMTs. Both axes are shown in the reference frame of XENONnT where $Z=\qty{0}{\centi \meter}$ corresponds to the position of the TPC gate electrode.}
    \label{fig_ambe_spatial}
\end{figure}

\isotope[241]{Am}Be calibration data was taken at three different locations close to the cryostat, at two different heights. 
A fourth position further away from the cryostat, close to the halfway position between the cryostat and NV walls, was used to further validate the NV response. 
The spatial distribution of the reconstructed NV events for one of the calibration source position close to the cryostat is shown in Fig.\,\ref{fig_ambe_spatial}.

An \isotope[241]{Am}Be source emits neutrons via the reaction $\isotope[9]{Be}(\mathrm{\alpha, n})\isotope[12]{C}$, ending either in the ground or first excited state of \isotope[12]{C}. 
This is an attractive feature for a calibration source as the emitted $\gamma$-ray from the first excited state can be used to cleanly identify and label neutron interactions in the NV or the TPC. 
This approach was also successfully used in the calibration of other water Cherenkov detectors \cite{sk_neutron_capture_water,sno_ncapture}.
The first excited state has an energy of {\ambenone}, and the neutron follows a continuous energy distribution with an average kinetic energy of about \qty{4.5}{MeV} \cite{ambe_scherzinger}. 
Additionally, the source can emit neutrons with lower energy via a neutron-breakup reaction \cite{ambe_geiger} 
\begin{equation}
\isotope[9]{Be}(\mathrm{\alpha, \alpha'})\isotope[9]{Be}^* \rightarrow \isotope[8]{Be} + \mathrm{n}\,.   
\end{equation} 
The branching ratio between the different decay modes and final states of \isotope[12]{C} depends on the kinetic energy of the $\mathrm{\alpha}$-particle impinging on \isotope[9]{Be} and therefore on the manufacturing details of the source \cite{ambe_scherzinger,ambe_geiger}.
In the literature, the branching ratio for the first excited state of \isotope[12]{C} is reported to be around \qty{60}{\percent} \cite{ambe_scherzinger} ranging down to about \qty{50}{\percent} \cite{ambe_Ito_2023}. 
Studies of the branching ratio performed during SR0 also indicate a branching ratio of about \qty{50}{\percent} for the source used in XENONnT \cite{thesis_daniel_wenz}. 
However, for the calibration of the NV performance, a precise knowledge of the branching ratio is not required, as only neutrons emitted in coincidence with a $\gamma$-ray are used in the analysis.
At the time of calibration, the source had a neutron rate of {\amberate} estimated based on the decay rate measured in \cite{XENON:2013nr_response} and the half-life of \isotope[241]{Am}.

\begin{figure}
\centering
\includegraphics[width=0.49\textwidth]{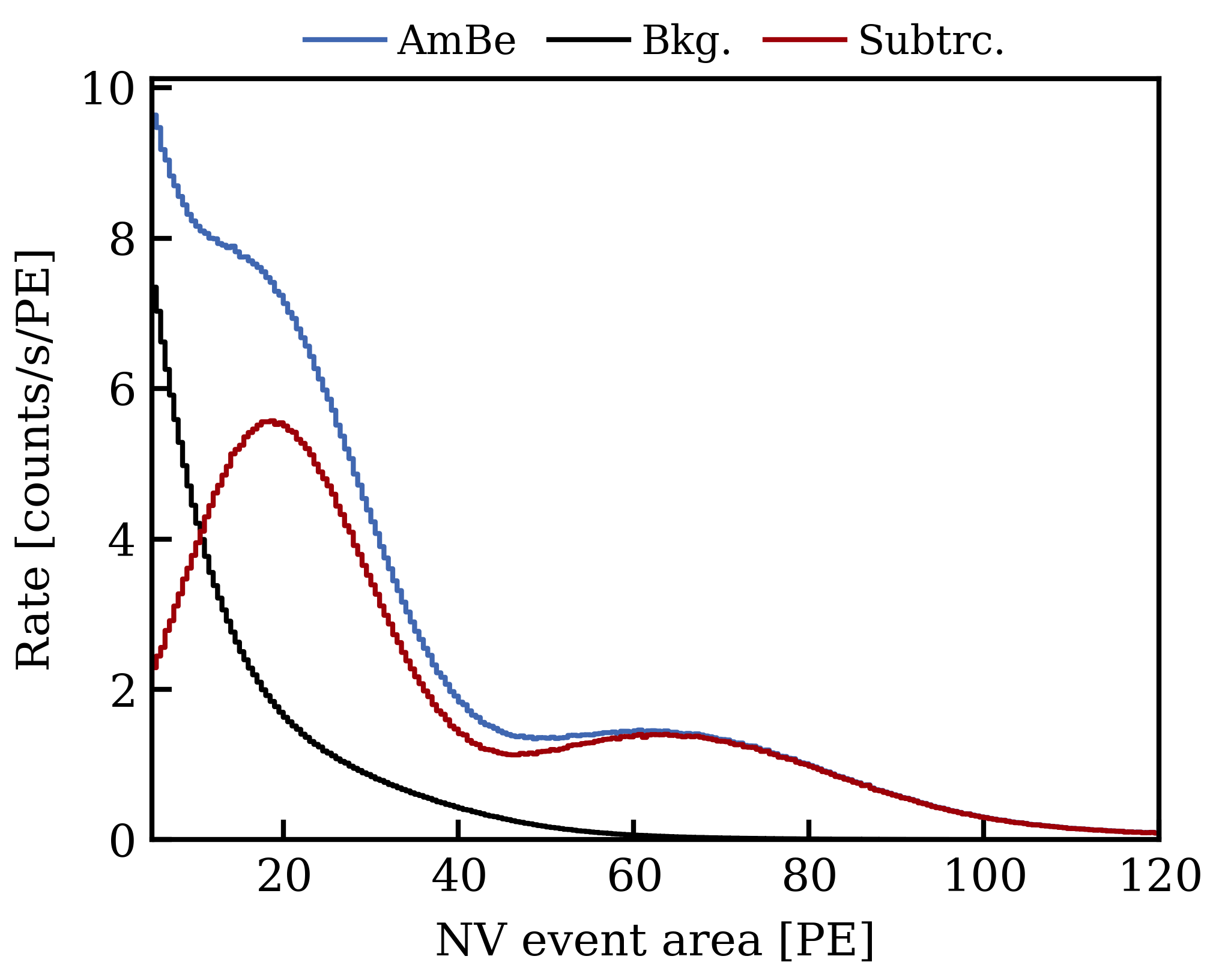}
\caption{\isotope[241]{Am}Be spectrum for a single calibration source position. The black and blue distributions show the background and calibration data, respectively. The red distribution shows the resulting binwise background-subtracted distribution. The first peak centered around \qty{20}{\PE} corresponds to the {\hydrogencaptureenergy} $\gamma$-ray of the neutron capture on hydrogen. The second peak around \qty{65}{\PE} orginates from the {\ambenone} $\gamma$-ray from the de-excitation of \isotope[12]{C}.}
\label{fig_ambe_spectrum}
\end{figure}

Fig. \ref{fig_ambe_spectrum} compares the \isotope[241]{Am}Be spectrum recorded by the NV with background data for one of the calibration source positions next to the cryostat.
A set of data-quality cuts based on the spatial distribution of the events and shape properties of the Cherenkov signal are applied.
Both the \qty{2.22}{\mev} neutron capture line on hydrogen and the {\ambenone} $\gamma$-line are clearly visible. 
The two signals are centered around \qty{20}{\PE} and \qty{65}{\PE}, respectively.
The tail of the distribution at large energies arises from neutron captures on other isotopes like \isotope[56]{Fe} in the stainless steel of the cryostat, which emits a cascade of $\gamma$-rays with a total energy up to \qty{7.64}{\mev}.

\subsection{Neutron veto detection efficiency}
\begin{figure}
    \centering
    \includegraphics[width=0.49\textwidth]{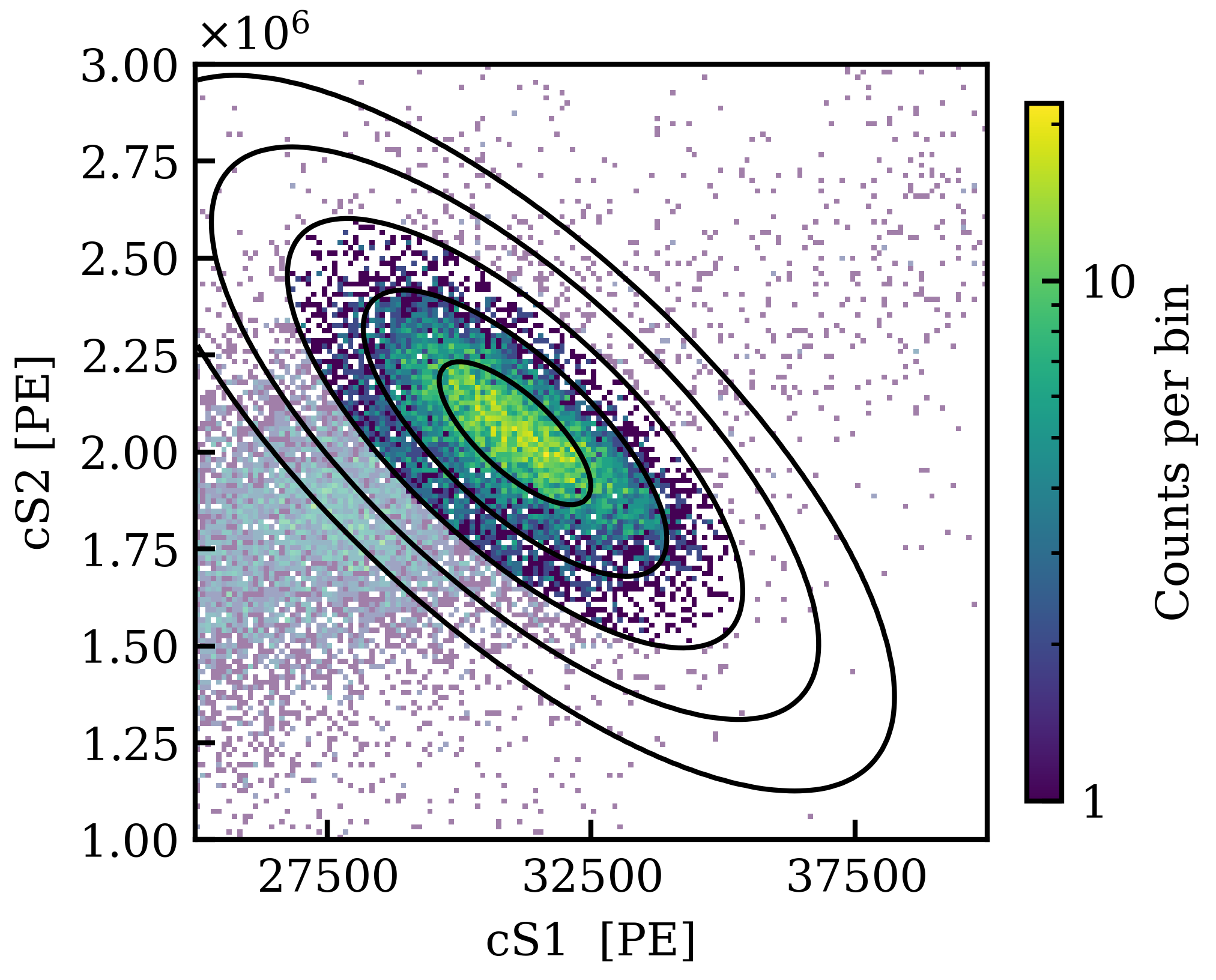}
    \caption{Distribution in the corrected S1-S2 space of the full absorption peak of the {\ambenone} $\gamma$-ray detected in the TPC. The black contours show the 1 to $5\,\sigma$ regions of the ellipse of the best fit. Only fully colored events within $3\,\sigma$ are selected.}
    \label{fig_full_energy_ellipse}
\end{figure}

The neutron detection efficiency is in general larger than the tagging efficiency, since neutrons are not required to interact inside the TPC before being captured by the surrounding water.
To count the number of emitted neutrons by the source, the {\ambenone} $\gamma$-ray recorded by the TPC is used as a trigger.
After applying a set of data-quality cuts, all {\ambenone} $\gamma$-ray signals within the $3\,\sigma$ contour of the full-energy ellipse in the corrected S1-S2 space were selected, as shown in Fig. \ref{fig_full_energy_ellipse}. 
As these signals are larger than any other background signal produced in the TPC by natural radioactivity, the selection can be considered background-free.

\begin{figure}
    \centering
    \includegraphics[width=0.49\textwidth]{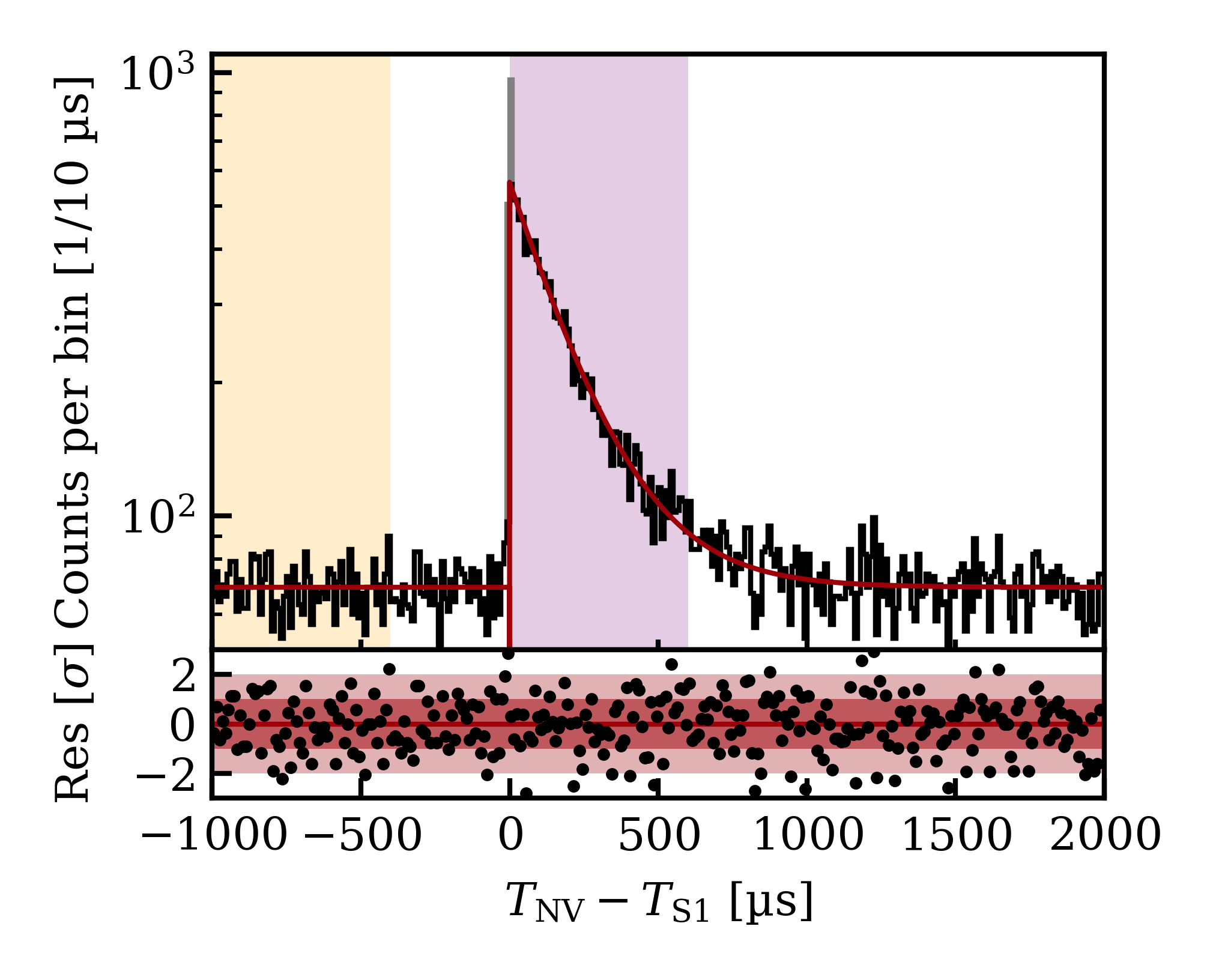}
    \caption{Distribution of the time difference between NV events and the {\ambenone} $\gamma$-ray S1 signals recorded in the TPC. The red line shows the best fit of the time distribution. The purple and orange areas indicate the region of interest (ROI) and background reference regions explained in the text. The black histogram shows a binned representation of the data, whereas the gray-coloured part indicates bins excluded from the fit region. The bottom panel shows the residuals of the best fit with the binned data.}
    \label{fig_time_distribution_detection_efficiency}
\end{figure}

To count the number of neutron capture events, NV events were searched in a wide coincidence window of [\qty{-1000}{\mus}, \qty{2000}{\mus}] between TPC and NV, given the capture time of neutrons in demineralized water being {\neutroncapturetimewater} \cite{sk_neutron_capture_water,sno_ncapture}. 
The resulting time distribution is shown in Fig. \ref{fig_time_distribution_detection_efficiency} and was fitted with an exponential distribution on top of a uniform background using an extended unbinned maximum-likelihood method. 
The region around [\qty{-0.1}{\micro \second}, \qty{0.1}{\micro \second}] was excluded from the fit to avoid any bias due to the {\ambenone} $\gamma$-ray itself. The decay constant of the exponential distribution $\tau_\text{C}$, which represents the neutron capture time, is \qty{194\pm4}{\mus}. This is slightly smaller than the expected value of {\neutroncapturetimewater} for a demineralized water target. 
The difference originates from radiative neutron capture on other isotopes like \isotope[56]{Fe}, which are part of the cryostat materials.
The observed effect becomes even stronger when estimating the neutron tagging efficiency in which by construction a neutron is required to enter the TPC. This was further confirmed by a systematic comparison of the decay constant and changes in the tail of the \isotope[241]{Am}Be energy distribution shown in Fig. \ref{fig_ambe_spectrum}, while varying the distance between the source and the TPC cryostat. 
More details can be found in \cite{thesis_daniel_wenz}, where a neutron capture time of \qty{202.1\pm0.2}{\mus} was measured when the source is positioned far away from the TPC cryostat, in very good agreement with the values measured by other water Cherenkov detectors \cite{sk_neutron_capture_water,sno_ncapture}.

The detection efficiency is computed as the number of capture events, determined by subtracting the number of events found inside the purple neutron capture region by the orange reference region depicted in Fig. \ref{fig_time_distribution_detection_efficiency}, over the number of selected {\ambenone} $\gamma$-ray trigger signals. 
This results in a detection efficiency of \qty{62\pm1}{\percent} for a \qty{250}{\micro \second} long capture time window, which becomes \qty{82\pm1}{\percent} for a longer \qty{600}{\micro \second} window. The uncertainty only includes statistical uncertainties. The estimated detection efficiency only represents a lower limit, as it is not corrected for the number of neutrons captured in the TPC which do not release energy in the NV. 
Despite this, it is the highest neutron-detection efficiency ever measured in a water Cherenkov detector \cite{sk_neutron_capture_water,sno_ncapture}.

\subsection{Neutron veto tagging efficiency}
In contrast to the detection efficiency, the tagging efficiency requires in addition that the emitted neutrons interact first inside the TPC producing an SSNR, before exiting and being captured in the NV. The SSNR events were selected by requiring a coincidence between the {\ambenone} $\gamma$-ray signal, this time detected in the NV, and the NR S1 signal detected in the TPC. This coincidence uses a much tighter window of $\tapprox$\qty{400}{\nano \second}, and selects well-reconstructed NR S1 signals with {\ambenrpurity} purity. Additional data-quality cuts are applied to the TPC events to select only SSNR events \cite{XENON:2024analysis_paper2}.

\begin{figure}
\centering
\includegraphics[width=0.49\textwidth]{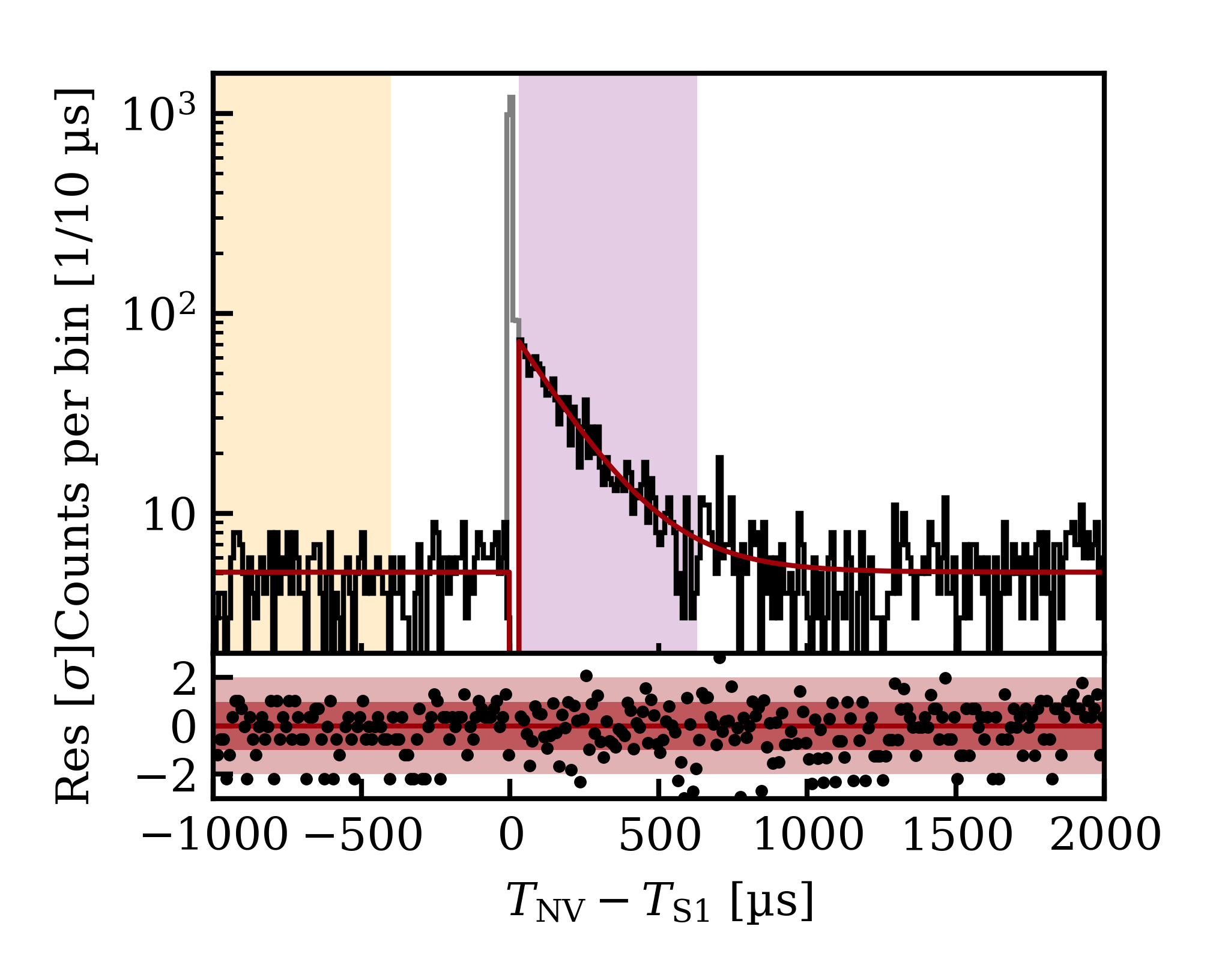}
\caption{Distribution of the time difference between NV events and the triggering single-scatter nuclear-recoil S1 in the TPC. The color code is the same as in Fig. \ref{fig_time_distribution_detection_efficiency}.}
\label{fig_time_distribution}  
\end{figure}

To estimate the tagging efficiency, the number of SSNR is compared against the number of neutron capture events detected by the NV. The procedure is similar to the one used for the detection efficiency. Neutron capture events are selected by requiring the same wide coincidence window of [\qty{-1000}{\mus}, \qty{2000}{\mus}] between NV events and the selected SSNR S1 signals. The relative time difference between the coincident events is shown in Fig. \ref{fig_time_distribution}. 

\begin{figure}
\centering
\includegraphics[width=0.49\textwidth]{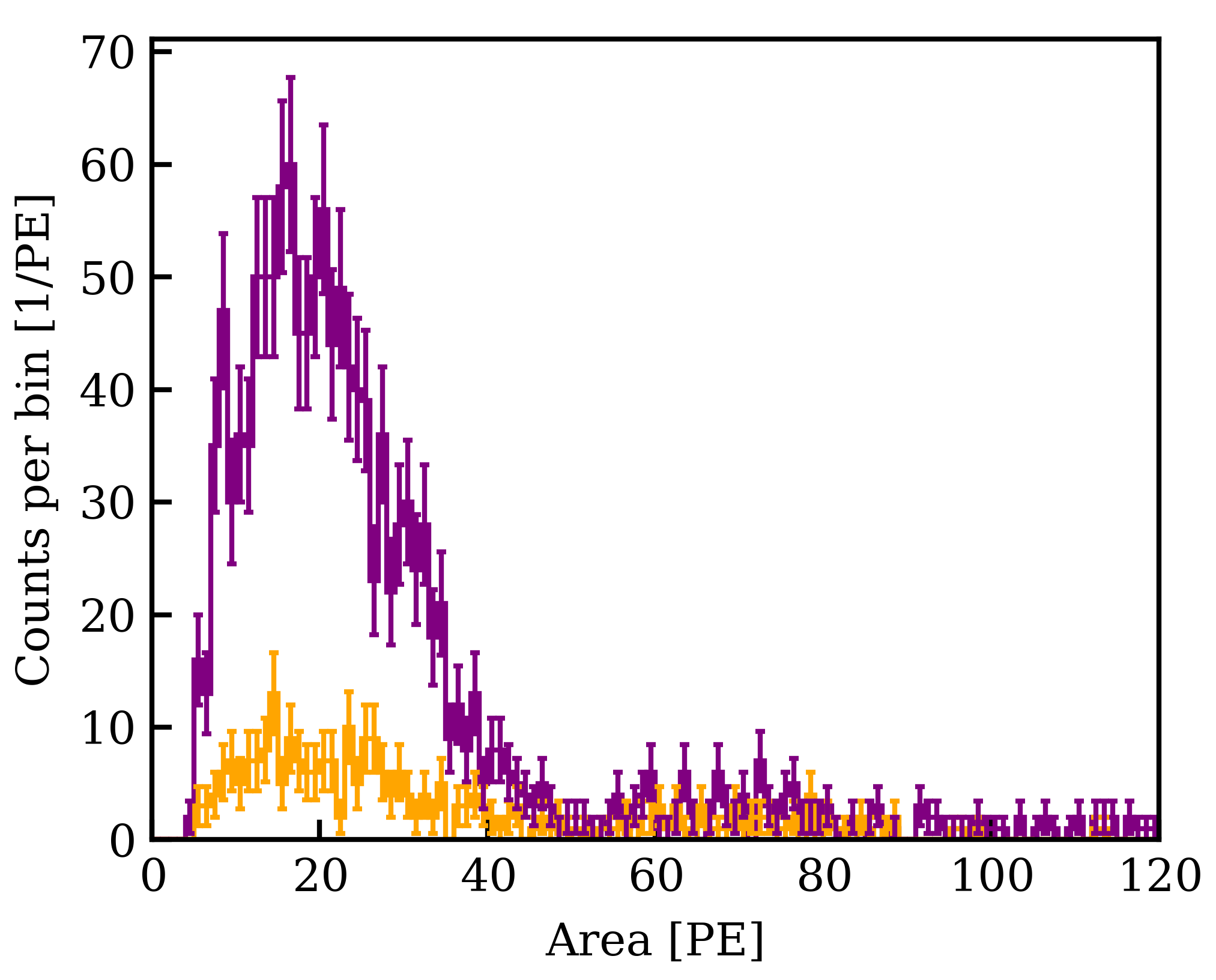}
\caption{Event area distribution for the signal (purple) and background reference region (orange), shown respectively in purple and orange in Fig. \ref{fig_time_distribution}. The error bars indicate the statistical uncertainties.}
\label{fig_events_nrss_roi_and_reference}  
\end{figure}

In contrast to Fig. \ref{fig_time_distribution_detection_efficiency}, the peak centered around zero is much more prominent. It indicates the {\ambenone} $\gamma$-ray events recorded by the NV, used to select the SSNR S1 signals.
The number of neutron capture signals is again computed by subtracting events inside the purple signal region [\qty{30}{\mus}, \qty{630}{\mus}), by events found in the orange reference region. 
The area distribution of both regions is shown in Fig. \ref{fig_events_nrss_roi_and_reference}.

The region between \qty{0}{\mus} and \qty{30}{\mus} is excluded from the computation of the number of capture events as it exhibits a higher NV background rate induced by the {\ambenone} $\gamma$-rays.
This includes not only the triggering {\ambenone} $\gamma$-rays themselves, but also accidental NV events caused by an increased rate of PMT afterpulses following the {\ambenone} signals.
Thus, to estimate the NV tagging efficiency, the background-subtracted number of neutron capture signals must be corrected for this chosen offset in the time window, as the veto window applied in the science data differs. 
The correction factor $\epsilon_\text{TW}$ can be estimated by taking the ratio between the fraction of the exponential distribution covered by the purple time window in Fig. \ref{fig_time_distribution} and the respective veto window applied in science data.
Thus, the correction is solely characterized by the decay constant $\tau_\text{C}$ of the exponential distribution.

Fig. \ref{fig_time_distribution} shows the best fit of the time distribution using the same model and fitting method as in Fig. \ref{fig_time_distribution_detection_efficiency}.
The decay constant of the exponential distribution $\tau_\text{C}$ is \qty{180\pm8}{\mus}.
The resulting time window correction factor $\epsilon_\text{TW}$ for a \qty{250}{\mus} and \qty{600}{\mus} long veto window is \qty{0.92\pm0.02}{} and \qty{1.18\pm0.01}{} respectively. 

Given that the \isotope[241]{Am}Be calibration was only conducted at a finite number of positions around the TPC cryostat, and that the neutron energy distribution of \isotope[241]{Am}Be and background neutrons are slightly different, an additional relative geometrical correction factor $\epsilon_\text{geo}$ was determined through Geant4 \cite{geant4} simulations. 
This correction factor takes into account the spatial distribution and yield of radiogenic neutrons from the detector materials \cite{xe1t_ana_paper_2}.
The correction is computed by taking the ratio between the simulated tagging efficiency for background neutrons with respect to the simulated tagging efficiency for \isotope[241]{Am}Be calibration.
While the latter is found to be slightly higher than the one observed in the data, both results are still consistent within their respective statistical uncertainties. From the simulation, a tagging efficiency of \qty{71\pm1}{\percent} is expected for an infinitely long tagging window, which compares to a value of \qty{70\pm3}{\percent} in data using a \qty{1200}{\micro \second} long tagging window.
The relative geometrical correction factor for radiogenic neutrons is $\epsilon_\text{geo} = (1.01\pm0.02)$, which is negligible compared to the correction of the time window. 

\begin{figure}
\centering
\includegraphics[width=0.49\textwidth]{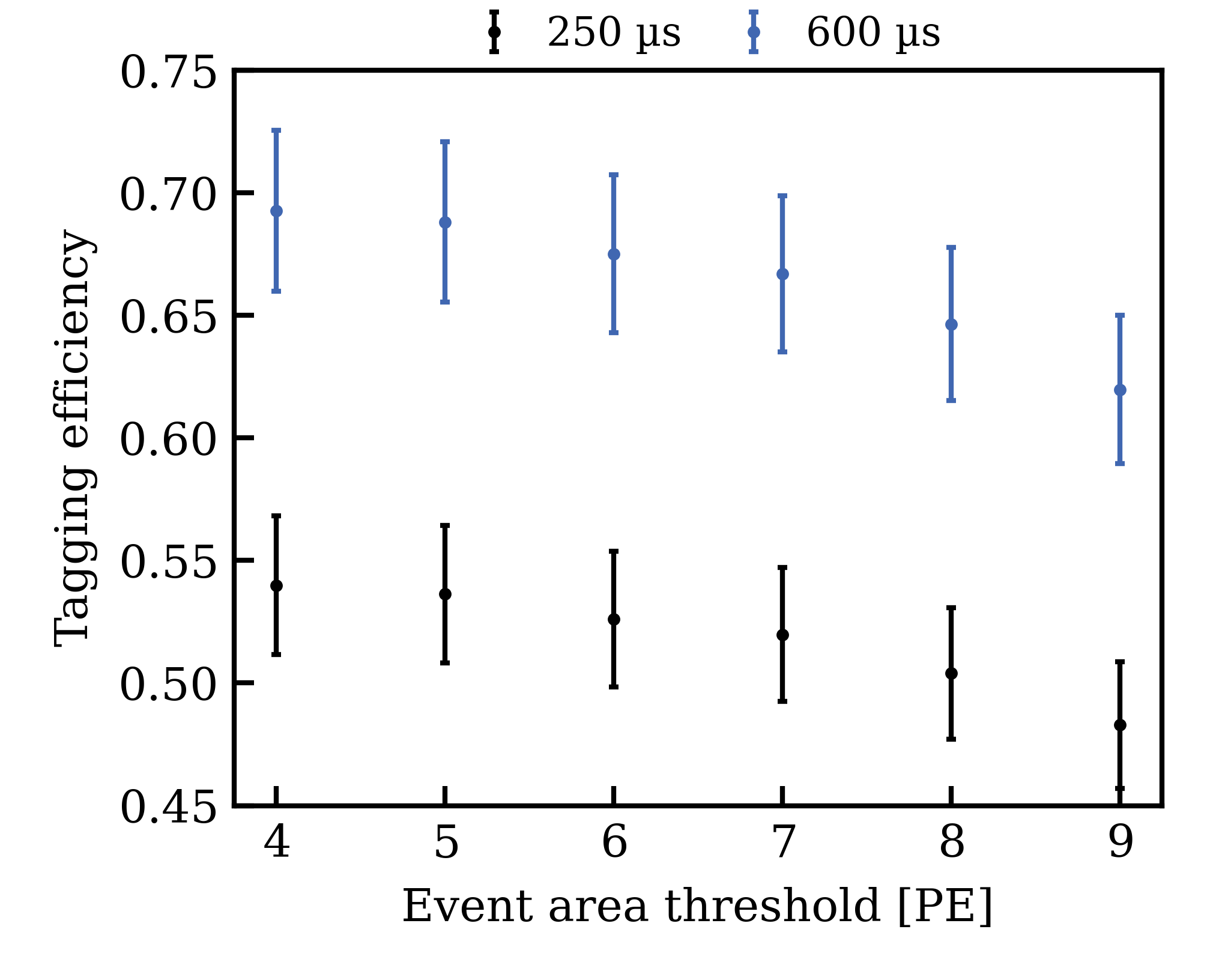}
\caption{Tagging efficiency of the NV in SR0, as a function of the threshold in event area. The black and blue data show the tagging efficiency for a \qty{250}{\mus} and \qty{600}{\mus} veto-window, respectively. In both cases, it was required that at least 5 PMTs contribute to an NV event. All the corrections explained in the text are applied.}
\label{fig_tagging_efficiency}  
\end{figure}

Fig. \ref{fig_tagging_efficiency} shows the final tagging efficiency after applying all corrections as a function of the event area threshold.
For SR0, a neutron-tagging window of \qty{250}{\micro \second} with an event area threshold of \qty{5}{\PE}, and a 5-fold PMT coincidence requirement was used, leading to a tagging efficiency of {\taggingefficiency}. The background rate in the NV with these selections is $\tapprox64$ Hz,  inducing a loss in the TPC live time of \qty{1.6}{\percent}. A \qty{600}{\micro \second} long window would increase the tagging efficiency to \qty{68\pm3}{\percent}, but was disfavored before unblinding due to the larger live time loss of \qty{3.8}{\percent}.

\section{Impact of the NV in SR0 and conclusion}
\label{sec:nv_in_sr0}
\begin{figure}
    \centering
    \includegraphics[width=0.49\textwidth]{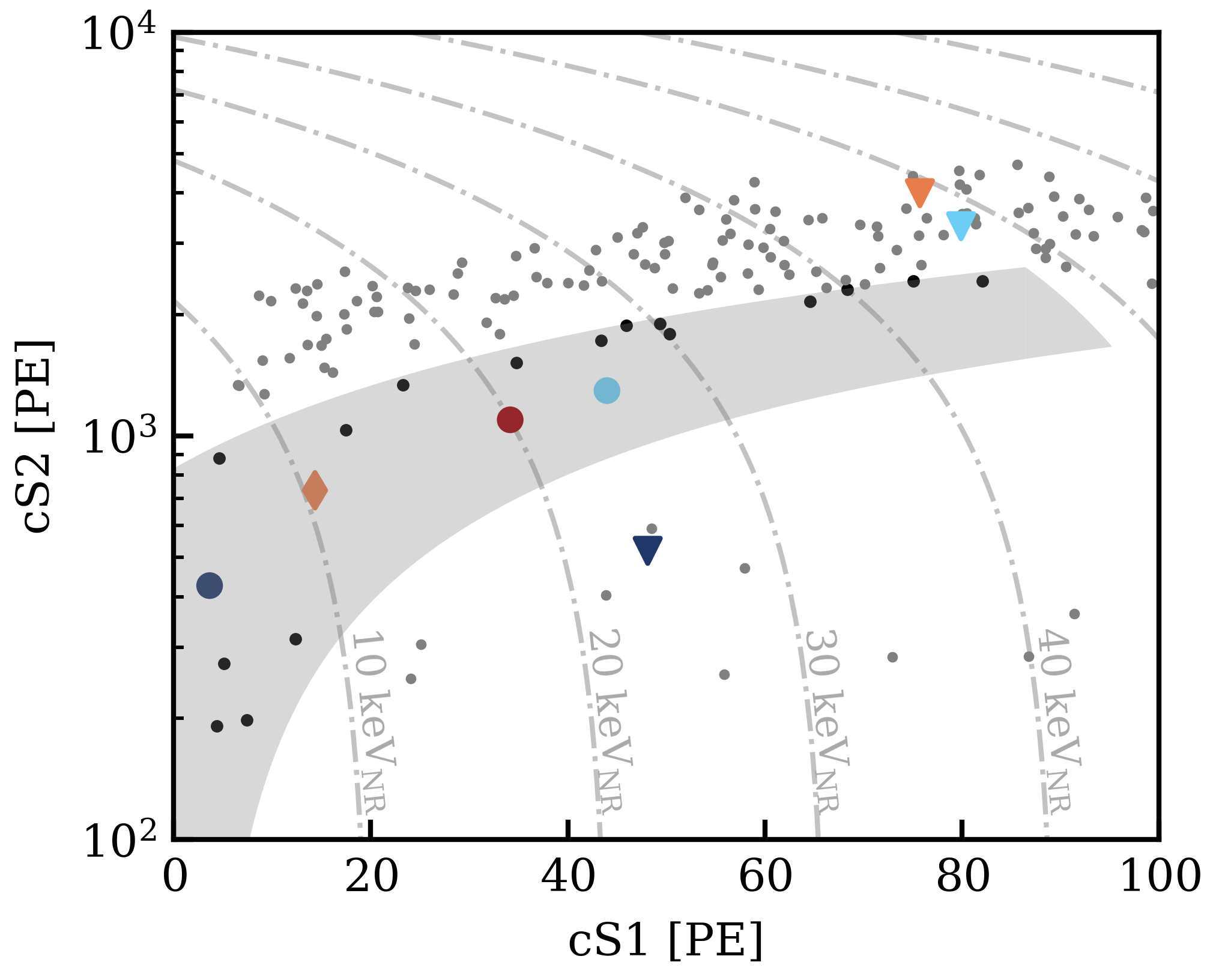}
    \caption{NV-tagged events in the WIMP ROI of the SR0 science data. The round marker represents the three tagged multi-scatter events, while the diamond indicates the tagged single-scatter signal. The color code of the markers encodes the event Id and it is the same as in Fig. \ref{fig_nv_tagged_sr0_events_nv_space} and \ref{fig_nv_tagged_sr0_events_spatial_reconstruction}. The gray-shaded region indicates the blinded region of SR0 where neutrons and WIMPs are expected. The gray dots indicate all events within the WIMP ROI, but outside of the blinded region, while black dots show all events inside. The dash-dotted lines indicate the iso-energy contours of NRs. Events tagged by the NV outside of the blinded region are indicated by colored triangles.}
    \label{fig_nv_tagged_sr0_events}
\end{figure}

\begin{figure}
    \centering
    \includegraphics[width=0.49\textwidth]{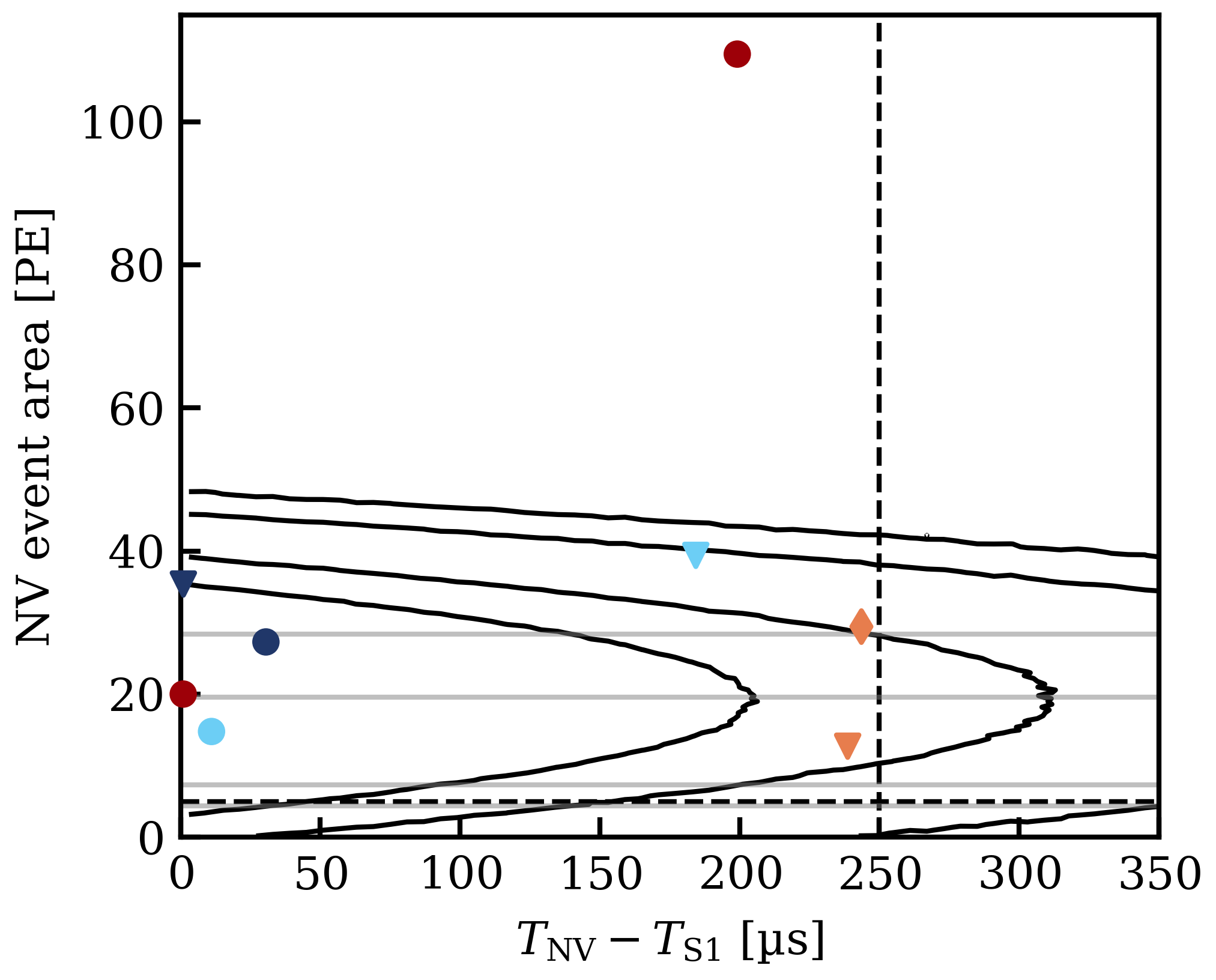}
    \caption{Area of NV tagged events as a function of time delay between S1 and NV signal. The dashed lines indicate the threshold and the tagging window length used in SR0. The black contours and gray horizontal lines indicate the \qty{50}{\percent}, \qty{70}{\percent}, \qty{90}{\percent} and \qty{95}{\percent} quantiles for NV neutron capture and background signals, respectively. The events are identified by the same color and marker as in Fig. \ref{fig_nv_tagged_sr0_events}.}
    \label{fig_nv_tagged_sr0_events_nv_space}
\end{figure}

\begin{figure}
    \centering
    \includegraphics[width=0.48\textwidth]{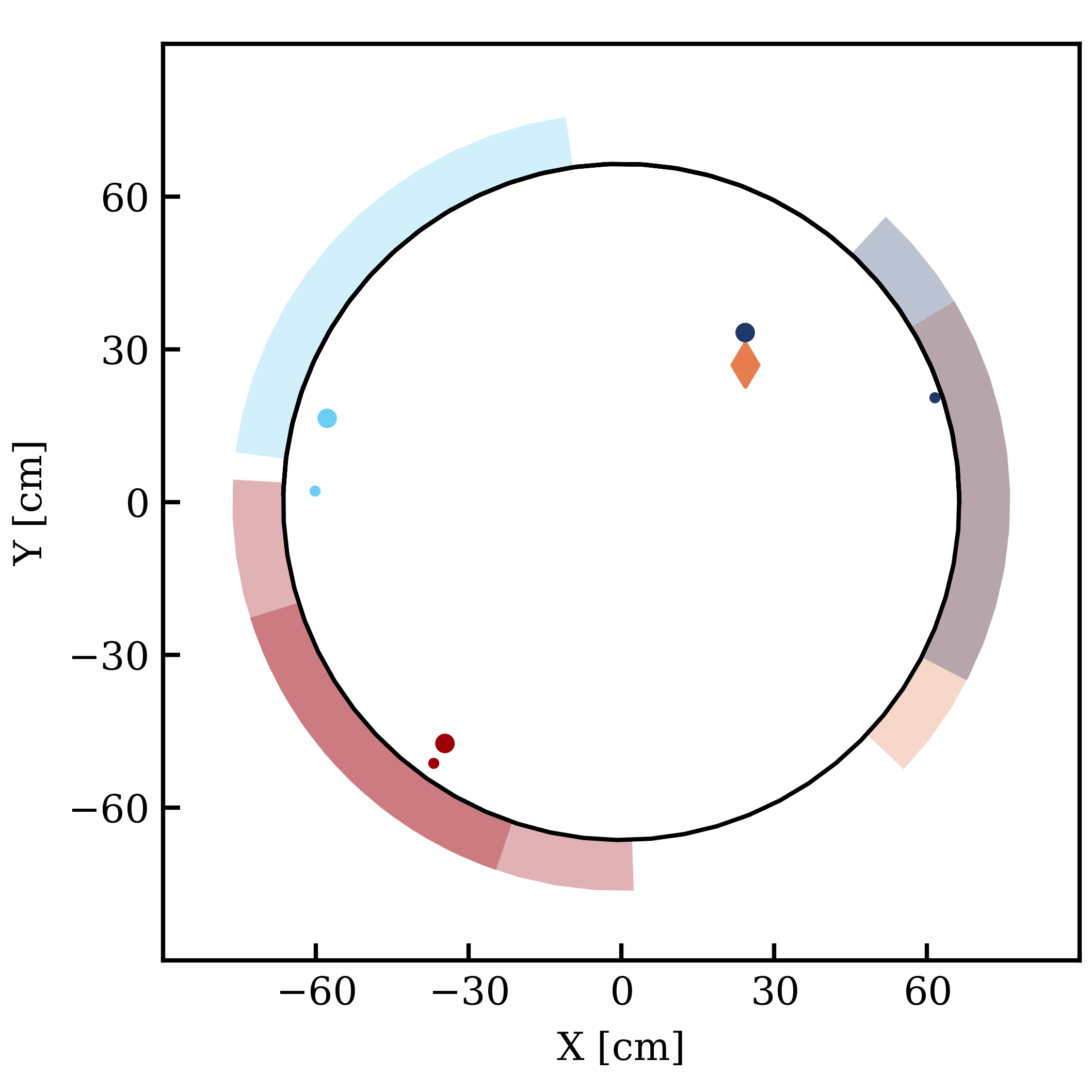}
    \caption{Spatial correlation between the TPC and NV events. The marker size indicates the position of the largest and second-largest S2 signal of the given event. The color-shaded wedges indicate the reconstructed azimuthal angle of the NV events. The angular coverage of the wedges corresponds to the average 1$\sigma$ contour of the \isotope[241]{Am}Be calibration distributions as shown in Fig. \ref{fig_ambe_spatial}. The red event shows two wedges, one for each NV event in Fig. \ref{fig_nv_tagged_sr0_events_nv_space}. The black circle indicates the boundary of the TPC. The events are identified by the same color and marker as in Fig. \ref{fig_nv_tagged_sr0_events}. Events outside the blinded WIMP region are omitted.}
    \label{fig_nv_tagged_sr0_events_spatial_reconstruction}
\end{figure}

In SR0, the WIMP ROI was defined between [\qty{0}{\PE}, \qty{100}{\PE}] in cS1 and [\qty{100}{\PE}, \qty{e4}{\PE}] in cS2, were the ``c'' indicates that the signals were corrected for detector dependent effects. In Fig. \ref{fig_nv_tagged_sr0_events}, all TPC events within the WIMP ROI are shown, highlighting those events tagged by the NV.
In total three multi-scatter and one single-scatter events were found in the blinded region, which is consistent with the expected multi-to-single-scatter ratio for neutron signals of about 2.2 obtained from MC simulations, and validated with the MS/SS ratio observed in the \isotope[241]{Am}Be calibration \cite{XENON:2022wimp,XENON:2024analysis_paper2}.
In addition, three events outside of the blinded region were tagged by the NV, which is consistent with the expectation from an accidental tagging of \qty{2.5\pm0.2}{} events given the background rate in the NV.
All four events within the blinded region show a high likelihood of being neutron-induced signals in terms of deposited energy inside the NV, and their time correlation with the TPC signals, as shown in Fig. \ref{fig_nv_tagged_sr0_events_nv_space}. 
One of the multi-scatter signals, identified by the red dot, is tagged by two different neutron veto events. 
The first event is within the \qty{50}{\percent} contour of neutron capture signals with a time difference between NV and TPC S1 signal of less than  \qty{1}{\micro \second}.
The second event, delayed by \qty{200}{\micro \second}, corresponds to a much larger energy deposit of several MeV which might originate from the deexcitation of a daughter nuclei if the tagged neutron was emitted during a spallation process.
Also, all four events show a spatial correlation between NV and TPC as shown in Fig. \ref{fig_nv_tagged_sr0_events_spatial_reconstruction}, which is an additional indication of neutron-induced signals.
Given these four tagged events, the neutron background expectation for SR0 was estimated to be $(1.1 ^{+0.6}_{-0.5})\,$events \cite{XENON:2022wimp,XENON:2024analysis_paper2}.

\vspace{0.5cm}

In conclusion, in this work we presented the commissioning and first results of the XENONnT water Cherenkov NV, the world's first water Cherenkov detector dedicated to tag neutrons through the low-energy emission of their capture in water. 
Throughout the first science run of \-XENONnT, the NV showed a robust PMT performance, constant background rate and stable water transparency, with variations inside $2\%$.

The efficiency for detecting neutrons was determined based on a coincidence technique between TPC and NV, exploiting the fact that neutrons emitted by an \isotope[241]{Am}Be source are often accompanied by a \ambenone $\gamma$-ray. 
The measured neutron-detection efficiency is \qty{82\pm1}{\percent}, the highest neutron-detection efficiency ever measured in a water Cherenkov detector.
This high detection efficiency also results in a high efficiency for tagging WIMP-like neutron signals inside the TPC. For the first science run of XENONnT, a reduced tagging window of \qty{250}{\mus} was used, setting the tagging efficiency to \qty{53\pm3}{\percent} and the overall livetime loss to \qty{1.6}{\percent}. 
A wider tagging window of \qty{600}{\mus} can increase the neutron tagging efficiency up to \qty{68\pm3}{\percent}, at a cost of a higher live time loss of \qty{3.8}{\percent}. 
One WIMP-like neutron signal was tagged by the NV, highlighting its importance for the WIMP dark matter search. 

Also in the second science run of XENONnT, started in 2022 the neutron veto was operated with demineralized water showing a similar excellent performance as during SR0. 
Afterwards, at the end of 2023, the water in the tank was doped with gadolinium, the element with the highest neutron capture cross-section. The Gd was added through Gd-Sulphate-Octa\-hydrate, at a concentration of 0.05\% in salt mass \cite{Marti:2019dof,Super-Kamiokande:2021the}. 
This improves the performance of neutron tagging thanks to the increased capture cross-section, the related decrease in capture time, and the increase in the total amount of energy released by the neutron capture on Gd, which is about \qty{8}{MeV} \cite{Gd_gamma_ray_model}.
A new plant dedicated to dissolving the Gd salt in water and continuously purifying the Gd-water solution has been installed and it is currently operating at LNGS. 
The Gd-water Cherenkov detector concept and the purification of the Gd-water solution, have been designed following the expertise developed in the EGADS and Super-Kamiokande experiments \cite{Marti:2019dof,Super-Kamiokande:2021the,SKSecondGdLoading}. 
First measurements of the Gd-doped NV performance show an improvement in 
tagging efficiency consistent with expectations, reducing the neutron background by a factor of $\tapprox2$ compared to SR0. These results will be described in a dedicated work, which is under preparation.


\begin{acknowledgements}

We gratefully acknowledge sup\-port from the National Science Foundation, Swiss National Science Foundation, German Ministry for Education and Research, Max Planck Gesellschaft, Deutsche Forschungsgemeinschaft, Helm\-holtz Association, Dutch Research Council (NWO), Fundacao para a Ciencia e Tecnologia, Weizmann Institute of Science, Binational Science Foundation, Région des Pays de la Loire, Knut and Alice Wallenberg Foundation, Kavli Foundation, JSPS Kakenhi and JST FOREST Program, and ERAN in Japan, Tsinghua University Initiative Scientific Research Program, DIM-ACAV\textsuperscript{+} R\'egion Ile-de-France, and Istituto Nazionale di Fisica Nucleare. This project has received funding/support from the European Union’s Horizon 2020 research and innovation program under the Marie Skłodowska-Curie grant agreement No 860881-HIDDeN. 

We gratefully acknowledge support for providing computing and data-processing resources of the Open Science Pool and the European Grid Initiative, in the following computing centers: the CNRS/IN2P3 (Lyon - France), the Dutch national e-infrastructure with the support of SURF Cooperative, the Nikhef Data-Processing Facility (Amsterdam - Netherlands), the INFN-CNAF (Bologna - Italy), the San Diego Supercomputer Center (San Diego - USA), and the Enrico Fermi Institute (Chicago - USA). We acknowledge the support of the Research Computing Center (RCC) at The University of Chicago for providing computing resources for data analysis.

We thank the INFN Laboratori Nazionali del Gran Sasso for hosting and supporting the XENON project.

\end{acknowledgements}


\bibliography{bibliography.bib}   

\end{document}